\documentclass[11pt]{article}
\usepackage{amsmath,amsfonts,amssymb,epsfig,color,psfrag}
\usepackage{rotating}
\begin{document}
\title{On P vs. NP, Geometric Complexity Theory, Explicit Proofs and
the Complexity Barrier}
\author{
Dedicated to Sri Ramakrishna \\ \\
Ketan D. Mulmuley 
 \\
The University of Chicago
\\  \\
http://ramakrishnadas.cs.uchicago.edu \\ \\
Technical Report, Computer Science Department,\\
The University of Chicago 
}

\maketitle

\newtheorem{prop}{Proposition}[section]
\newtheorem{claim}[prop]{Claim}
\newtheorem{goal}[prop]{Goal}
\newtheorem{theorem}[prop]{Theorem}
\newtheorem{metathesis}[prop]{Metathesis}
\newtheorem{mainpoint}{Main Point}
\newtheorem{hypo}[prop]{Hypothesis}
\newtheorem{guess}[prop]{Guess}
\newtheorem{problem}[prop]{Problem}
\newtheorem{axiom}[prop]{Axiom}
\newtheorem{question}[prop]{Question}
\newtheorem{remark}[prop]{Remark}
\newtheorem{lemma}[prop]{Lemma}
\newtheorem{claimedlemma}[prop]{Claimed Lemma}
\newtheorem{claimedtheorem}[prop]{Claimed Theorem}
\newtheorem{cor}[prop]{Corollary}
\newtheorem{defn}[prop]{Definition}
\newtheorem{ex}[prop]{Example}
\newtheorem{conj}[prop]{Conjecture}
\newtheorem{obs}[prop]{Observation}
\newtheorem{phyp}[prop]{Positivity Hypothesis}
\newcommand{\bitlength}[1]{\langle #1 \rangle}
\newcommand{\ca}[1]{{\cal #1}}
\newcommand{\floor}[1]{{\lfloor #1 \rfloor}}
\newcommand{\ceil}[1]{{\lceil #1 \rceil}}
\newcommand{\gt}[1]{{\langle  #1 |}}
\newcommand{\C}{\mathbb{C}}
\newcommand{\N}{\mathbb{N}}
\newcommand{\R}{\mathbb{R}}
\newcommand{\Z}{\mathbb{Z}}
\newcommand{\frcgc}[5]{\left(\begin{array}{ll} #1 &  \\ #2 & | #4 \\ #3 & | #5
\end{array}\right)}

\newcommand{\cgc}[6]{\left(\begin{array}{ll} #1 ;& \quad #3\\ #2 ; & \quad #4
\end{array}\right| \left. \begin{array}{l} #5 \\ #6 \end{array} \right)}

\newcommand{\wigner}[6]
{\left(\begin{array}{ll} #1 ;& \quad #3\\ #2 ; & \quad #4
\end{array}\right| \left. \begin{array}{l} #5 \\ #6 \end{array} \right)}

\newcommand{\rcgc}[9]{\left(\begin{array}{ll} #1 & \quad #4\\ #2  & \quad #5
\\ #3 &\quad #6
\end{array}\right| \left. \begin{array}{l} #7 \\ #8 \\#9 \end{array} \right)}

\newcommand{\srcgc}[4]{\left(\begin{array}{ll} #1 & \\ #2 & | #4  \\ #3 & |
\end{array}\right)}

\newcommand{\arr}[2]{\left(\begin{array}{l} #1 \\ #2   \end{array} \right)}
\newcommand{\unshuffle}[1]{\langle #1 \rangle}
\newcommand{\ignore}[1]{}
\newcommand{\f}[2]{{\frac {#1} {#2}}}
\newcommand{\tableau}[5]{
\begin{array}{ccc} 
#1 & #2  &#3 \\
#4 & #5 
\end{array}}
\newcommand{\embed}[1]{{#1}^\phi}
\newcommand{\stab}{{\mbox {stab}}}
\newcommand{\perm}{{\mbox {perm}}}
\newcommand{\trace}{{\mbox {trace}}}
\newcommand{\polylog}{{\mbox {polylog}}}
\newcommand{\sign}{{\mbox {sign}}}
\newcommand{\proj}{{\mbox {Proj}}}
\newcommand{\poly}{{\mbox {poly}}}
\newcommand{\std}{{\mbox {std}}}
\newcommand{\m}{\mbox}
\newcommand{\formula}{{\mbox {Formula}}}
\newcommand{\circuit}{{\mbox {Circuit}}}
\newcommand{\sgn}{{\mbox {sgn}}}
\newcommand{\core}{{\mbox {core}}}
\newcommand{\orbit}{{\mbox {orbit}}}
\newcommand{\cycle}{{\mbox {cycle}}}
\newcommand{\ideal}{{\mbox {ideal}}}
\newcommand{\qed}{{\mbox {Q.E.D.}}}
\newcommand{\proof}{\noindent {\em Proof: }}
\newcommand{\weight}{{\mbox {wt}}}
\newcommand{\tab}{{\mbox {Tab}}}
\newcommand{\level}{{\mbox {level}}}
\newcommand{\vol}{{\mbox {vol}}}
\newcommand{\vect}{{\mbox {Vect}}}
\newcommand{\val}{{\mbox {wt}}}
\newcommand{\sym}{{\mbox {Sym}}}
\newcommand{\convex}{{\mbox {convex}}}
\newcommand{\spec}{{\mbox {spec}}}
\newcommand{\strong}{{\mbox {strong}}}
\newcommand{\adm}{{\mbox {Adm}}}
\newcommand{\eval}{{\mbox {eval}}}
\newcommand{\for}{{\quad \mbox {for}\ }}
\newcommand{\Q}{Q}
\newcommand{\mand}{{\quad \mbox {and}\ }}
\newcommand{\invlim}{{\mbox {lim}_\leftarrow}}
\newcommand{\directlim}{{\mbox {lim}_\rightarrow}}
\newcommand{\sformal}{{\cal S}^{\mbox f}}
\newcommand{\vformal}{{\cal V}^{\mbox f}}
\newcommand{\crystal}{\mbox{crystal}}
\newcommand{\conje}{\mbox{\bf Conj}}
\newcommand{\graph}{\mbox{graph}}
\newcommand{\ind}{\mbox{index}}

\newcommand{\rank}{\mbox{rank}}
\newcommand{\id}{\mbox{id}}
\newcommand{\str}{\mbox{string}}
\newcommand{\RSK}{\mbox{RSK}}
\newcommand{\wt}{\mbox{wt}}
\setlength{\unitlength}{.75in}

\subsection*{Abstract}
Geometric complexity
theory (GCT) is an approach to the $P$ vs. $NP$ and related problems.
This article gives its  complexity theoretic overview  without 
assuming any background in  algebraic geometry or representation theory. 
\ignore{It describes the main barrier towards these problems called the 
{\em feasibility barrier},  the basic strategy 
of GCT for crossing it called the {\em flip}, 
which is to go for {\em explicit
proofs}, the main results of GCT that give a conjectural scheme to
construct such  proofs for the $P$ vs. $NP$ and related problems, 
and two concrete lower bound results of GCT which support this
strategy. By an explicit proof  we mean a proof 
that shows existence of  proof certificates 
of hardness for the $NP$-complete function under consideration 
that are short (of polynomial size) and easy to verify (in polynomial time). 
The two concrete lower bounds of GCT based on 
this strategy and its local form  are the $P\not = NC$ result without
bit operations and a mathematical form of the $\#P \not = NC$ conjecture
in characteristic zero.}

\section{Introduction} \label{sintro}
\begin{figure} 
\begin{center}
\epsfig{file=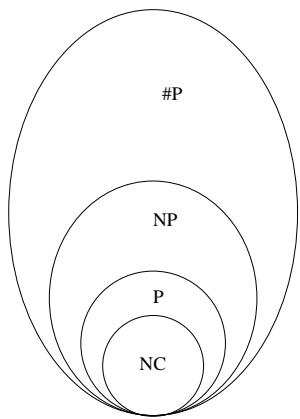, scale=.6}
\end{center}
      \caption{Complexity Classes (no proper containments are known)}
      \label{fig:complexityclasses}
\end{figure}

Geometric complexity
theory (GCT) is an approach to  the $P$ vs. $NP$ 
\cite{cook,karp,levin}, permanent vs. 
determinant \cite{valiant} and related problems
(Figure~\ref{fig:complexityclasses}) suggested 
in a series of  articles we call GCTlocal \cite{GCTlocal},
GCT1-8 \cite{GCT1}-\cite{GCT8}, and GCTflip
\cite{GCTflip}.
There seems to be a fundamental root difficulty, 
which we call the {\em complexity
barrier}, that must be overcome by any proof technique that could settle 
these problems.  Not surprisingly this barrier turns out to be 
extremely formidable.
All the mathematical effort in   GCT in a sense 
goes towards crossing this barrier. In fact, even formalization
of this barrier turns out to be quite nonelementary and cannot be done at 
present without algebraic geometry and representation theory.
The main result of GCT [GCT6] provides such a formalization and 
a mathematical programme for crossing the barrier, which has been 
partially implemented in  [GCT6,7,8,11]--this turns 
out to be quite a nonelementary affair. 
This paper provides an informal  exposition of the 
complexity barrier to explain why it may  be the root difficulty 
of the $P$ vs. $NP$ problem and what makes it so formidable
so that the need for  these nonelementary techniques 
that enter into GCT to tackle this  barrier becomes clear at a high level.

But before we turn to this barrier, let us first see why the earlier 
known barriers such as the relativization \cite{solovay}, natural-proof
\cite{rudich}, algebraic degree [GCTlocal] and  algebrization \cite{aaronson}
barriers do not really capture the 
main source of difficulties that needs to be overcome by an 
eventual proof.
This became clear after these barriers were bypassed in the early papers of 
GCT [GCTlocal,GCT1,GCT2]. Let us now explain how these barriers 
were bypassed in these early papers, and then 
explain why the root difficulty still remained after bypassing them.

Let us begin with [GCTlocal] 
that  proves a special case of the $P\not = NC$ conjecture  called 
{\em the $P\not = NC$ result  without bit operations}.
This may be considered to be the first concrete lower bound result of GCT.
It was proved using a weaker form of GCT's strategy  for crossing
the complexity barrier (discussed below) 
much before it was formalized in [GCTflip,GCT6]. 
The result  says that
the maxflow problem cannot be solved  in the PRAM model without bit operations
in polylogarithmic time using polynomially many processors.
At present this may be the only superpolynomial lower bound 
that is a special case of any of the fundamental separation problems in
Figure~\ref{fig:complexityclasses}
and  holds unconditionally in a natural and realistic 
model of computation that is  heavily used  in practice.
Furthermore, as explained in Section~\ref{swithoutbit}, 
its  proof technique is fundamentally different from 
the proof techniques to which the relativization 
and natural-proof barriers apply. 

But that is not enough. Because GCTlocal  pointed out 
a third barrier in complexity theory, which we shall call 
the {\em algebraic degree barrier}, but which was not called a barrier 
there for the reasons explained below. 
This basically says that any low degree or degree based proof technique, 
such as the one there,
which associates with a computation algebraic objects 
and then reasons solely on the basis of the degrees of those 
objects will not work for unrestricted 
fundamental separation problems. 
As far as we can see, 
the  recent algebrization barrier \cite{aaronson} also 
suggests   something similar \footnote{But on the basis of lower bounds 
which may not be as closely related with the fundamental separation
problems in Figure~\ref{fig:complexityclasses} as 
the $P\not = NC$ result without bit operations in GCTlocal.}.
Hence it is not treated  separately in this article.

GCTlocal also suggested 
an idea for crossing the algebraic degree barrier:
namely, associate with complexity 
classes algebraic varieties with group actions that capture the symmetries
of computation and then reason
on the basis of the deeper representation theoretic 
structure of these varieties rather than just their degrees.

This led to the investigation in GCT1 and 2, which suggested 
an approach to the fundamental lower bound problems via algebraic geometry
and representation theory. This  approach 
bypasses the relativization, natural proof and algebraic degree barriers
simultaneously as explained in Sections~\ref{scharsym} and \ref{smathform}. 
In contrast, it seems  \cite{aaronson} that the earlier known  techniques
for proving lower bounds in complexity theory
may not bypass  these  well studied barriers simultaneously.

Unfortunately after  GCT1 and 2, wherein the earlier known 
barriers were thus bypassed,
it became clear that  the root 
difficulty of the $P$ vs. $NP$ problem still remained after bypassing
these barriers  and the  main battle 
on the $P$ vs. $NP$ problem  really begins only after bypassing them.

In retrospect, this should not be surprising. 
The relativization,  natural-proof,
and algebraic-degree  barriers cannot be  expected to 
capture the root difficulty of the $P$ vs. $NP$ problem. 

To see why, 
let us begin with the algebraic degree barrier in [GCTlocal].
It applies to the low degree techniques of the kind used in that
paper.  All these techniques are in one way or
the other related to the fundamental Bezout's theorem in algebraic geometry
and the notion of a degree of a polynomial
(we also include the low degree techniques used in other areas of 
complexity theory in this category). Though this is a nontrivial area of
mathematics, it is still very small compared to the whole of mathematics.
Hence, we can say that this  barrier is {\em local} 
in the sense that it applies to a very 
restricted class of techniques in mathematics. 
If a proof technique does not belong to this class, the barrier 
is simply bypassed, i.e., is not applicable. 
In this sense, it  is not really a barrier--if by a barrier
we mean a conceptual obstacle 
that any eventual proof of the $P\not = NP$ conjecture would need to 
overcome regardless of which areas of mathematics it goes through--but 
rather a trap  that a proof technique has to avoid.
This is why it  was not  called a barrier in GCTlocal.

We can also say 
that the algebraic degree  barrier is primarily {\em mathematical} 
in the sense that low degree 
is a purely a mathematical notion, there is no complexity theory in it,
and though it  is very useful in complexity theory, this barrier is really a 
statement about the restricted nature   of these mathematical 
techniques rather than about the complexity theoretic difficulty 
of  the $P$ vs. $NP$ problem.

In a similar way, we may say that the relativization and natural proof 
barriers are also local and mathematical.  Mathematical  because the natural 
proof barrier is generally regarded as applying to proof techniques that 
work for most functions and most is a mathematical notion, 
so also the notion of relativization.

Now we can explain why  local mathematical 
barriers cannot be expected to
capture the root difficulty of the $P$ vs. $NP$ problem.
Because this problem is a {\em universal complexity-theoretic} 
statement about all of mathematics.
Namely, if this conjecture fails to hold,
every theorem in mathematics that has a proof of reasonable 
length  can be proved  efficiently by a computer in 
time polynomial in the length of its shortest proof. No other known
conjecture in mathematics has such stunning mathematical sweep. 
Given the universal complexity theoretic 
nature of this problem, it then stands to reason that
the main  barrier towards this problem--the root cause of its 
difficulty--should also be {\em universal} and {\em complexity theoretic},
where by a universal barrier we
mean a barrier that every approach has to cross regardless of which
areas of mathematics it goes through. 
In other words, a local barrier is like 
a guiding post which tells 
us what must not be done, and hence must  be bypassed,  whereas 
a universal barrier cannot be bypassed no matter what, because it tells us
what must be done, i.e., it has to be crossed.
A local barrier only  rules out a 
restricted  area of mathematics. 
Regardless of how many restricted   areas of mathematics we rule 
out in this way, that still rules  out only a 
small subarea of mathematics, tells us a few things that we should not do,
but gives us  essentially no idea about what to do
given that the $P$ vs. $NP$ problem is a statement about all of mathematics. 
At some point we have to turn our attention from what is not to be done
to  what is to be done
and identify the root universal and complexity theoretic 
difficulty of this problem; i.e., 
a universal complexity theoretic barrier.

The complexity  barrier described in this article 
seems to be this universal complexity theoretic barrier.
Crossing this barrier basically means   answering {\em formally}
the folklore question: why should the
$P$ vs. $NP$ problem   not preclude its own proof by standing 
in the way of any  approach, and hence, why should any given approach 
be feasible even theoretically? The key phrase here is answering formally. 
This  means:
(1) fix an appropriate  proof strategy (approach) and
formalize the question by defining precisely and
formally what
theoretically feasibility of this fixed proof strategy
means, and (2) then answer the formalized question by proving that this
proof strategy is indeed theoretically feasible as per this definition.
In other words, formalization of the informal folklore question as in step (1)
is itself a  part of the challenge. If this is carried out, we say that 
the complexity barrier has been formalized in this approach. 
The $P\not = NP$ conjecture seems so severe that for most proof strategies
it may not  even be possible to carry out step (1). That is, formal
definition theoretical feasibility may be possible 
for only some exceptional proof strategies. This is why just 
formalization of the complexity  barrier seems  to be such a challenge.

The most obvious and natural abstract strategy to cross the complexity 
barrier, called the {\em flip} [GCT6,GCTflip], 
is suggested by the $P$ vs. $NP$ problem itself. It is to 
go for its    {\em explicit proof}. By this we
mean a proof  that shows existence of  proof certificates
of hardness for an NP-complete function $f(X)=f(x_1,\ldots,x_n)$ 
that are {\em short} of $\poly(n)$ size and  {\em easy to verify} 
in $\poly(n)$ time; i.e., the problem of verification 
of the proof certificates belongs to $P$ and hence is theoretically feasible.
In other words, the flip seeks proof certificates 
for nonmembership of $f(X)$ in $P$ akin to those for its membership in $NP$.
Abstractly, GCT is any geometric 
approach to the $P$ vs. $NP$ and related problems based on the flip; i.e., 
on construction of  explicit proofs.
Such explicit 
proofs based on the specific 
proof certificates  defined in GCT1,2 and 6, called 
{\em geometric  obstructions}, 
exist as per the {\em explicit proof conjecture} [GCT6].
The main result of  GCT  is a complete formalization [GCT6] of the 
complexity barrier (in this specific approach) and a
mathematical programme 
[GCT6,7,8,11] to  cross the formalized complexity  barrier via the flip
for the $P$ vs. $NP$ and related problems. 
\ignore{It is supported by 
two concrete lower bounds of GCT based on  the flip and its local form: namely,
{\em the  $P \not = NC$ result without bit operations} [GCTlocal] 
and {\em a mathematical  form of the $\#P\not = NC$ conjecture in 
characteristic zero} [GCT1,2,6] discussed above.}

This article  gives a complexity theoretic 
overview  of GCT outlining these developments. 

This also allows us to  place  GCT in the landscape of all possible 
approaches to cross this main barrier in complexity theory.
For that, we define  GCT abstractly 
as any geometric approach based on the flip 
towards the $P$ vs. $NP$ and related 
problems, where geometric means based on
the  symmetries of a judiciously chosen $f(X)$. The present formal
GCT is one such approach. 
Though the flip   is the most obvious and natural 
way   to cross the complexity barrier, 
paradoxically it also  turns out to be  very counterintuitive and hard to
believe at the same time.
The formalization of the complexity barrier mentioned above 
basically provides a concrete mathematical program to implement the flip and 
says that the flip
should be feasible however counterintuitive it may seem.
The main goal of this article is to justify this from the complexity
theoretic perspective.
Two concrete lower bounds of GCT mentioned above play a crucial role in
this justification.

The complementary article \cite{GCTriemann} gives a mathematical overview 
of GCT based on the IAS lectures 
without getting into the meta
issues concerning  barriers. The reader who wishes to 
get a concrete mathematical picture of the approach 
before worrying about such issues may wish to read that article first. On the 
other hand, the reader who wishes to know the root difficulty of the $P$ 
vs. $NP$ and related problems and the need for  nonelementary 
techniques  before getting into any mathematics 
may wish to read this article first. We leave the choice to the readers.
See also the recent article \cite{landsberg} for 
an external mathematical  review of GCT1 and 2.

The rest of this article is organized as follows. The complexity barrier
is discussed in Section~\ref{scomplexity},  high level statement of 
the main result of GCT--a formalization of the 
complexity barrier--in Section~\ref{sGCT}.
Abstract  flip is described in Section~\ref{sflip},
the $P\not = NC$ result without bit operations
 in Section~\ref{swithoutbit},
the meaning of the phrase geometric in Section~\ref{scharsym},
why GCT bypasses the earlier known local barriers simultaneously 
in Section~\ref{smathform}, elaboration of the 
main result  in 
Sections~\ref{sgct6} and \ref{sexpli},  and finally the question--is the
complexity barrier really so formidable?--in Section~\ref{sformidable}. 
Sections~\ref{scharsym}-\ref{sexpli} constitute
an extended abstract of Chapter 1 in \cite{GCTriemann}.

\noindent {\em Remark:} 
Natural in GCT means explicit unlike in  \cite{rudich} where it means
probabilistic. The meaning should be clear from the context.

\noindent {Acknowledgements:} The author is deeply grateful to Janos Simon
without whom this paper would not exist.
The paper grew out of a series of discussions on GCT with 
him over the last year.
The goal was to communicate to him the root difficulty that GCT is trying to 
overcome and its  basic plan for doing so. That root difficulty is the
complexity barrier described here. This paper  basically consists of
answers to his tough and insightful questions. 
The author is also grateful to  Scott Aaronson, 
Laci Babai, Josh Grochow, Russell Impagliazzo,
Hari Narayanan, Partha Niyogi, Sasha Razborov, and
Avi Wigderson for helpful discussions.

\section{Complexity barrier} \label{scomplexity}
In this section we describe the complexity barrier.

Towards that end,    let us 
fix an  $NP$-complete function $f(X)=f(x_1,\ldots,x_n)$, say SAT. 
The goal of the nonuniform $P$ vs. $NP$ problem  is to show that 
there does not exist a circuit $C=C(X)$ of size 
$m=\poly(n)$  that computes $f(X)$, $n\rightarrow \infty$. 
Symbolically, let  $f_C(X)$ denote the function computed by $C$. 
Then we want to prove that
\begin{equation} \label{eqiohp}
\forall n, m=\poly(n) \ \forall C \exists x : f(X)\not = f_C(X). 
\end{equation}

\begin{figure} 
\begin{center}
\epsfig{file=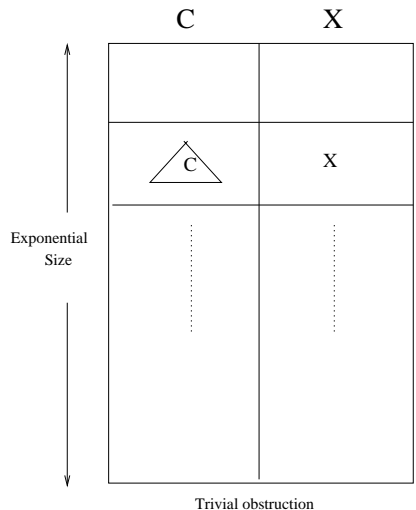, scale=.5}
\end{center}
      \caption{Trivial obstruction}
      \label{fig:trivialobs}
\end{figure}

Equivalently, the goal is to prove:

\noindent {\bf (IOH)}: For every large enough $n$, and $m=\poly(n)$,
there  exists a {\em trivial obstruction
(i.e. a ``proof-certificate'' of hardness)} to efficient computation of
$f(X)$. Here by a trivial obstruction we mean
a table (cf. Figure~\ref{fig:trivialobs}) that lists for every small 
$C$  a counterexample  $X$ such that $f(X)\not = f_C(X)$.

The number of rows of this table is equal
to the number of circuits $C$'s of size $m=\poly(n)$.
Thus the size of this table is 
exponential; i.e., $2^{O(\poly(n))}$. 
The time to verify whether a given table is a trivial obstruction
is also exponential,
and so is the time of the obvious algorithm 
to decide if such a table exists for given $n$ and $m$. 
From the complexity theoretic viewpoint, this is an infeasible 
(inefficient) task. So
we call  this trivial, brute force 
strategy of proving the nonuniform $P$ vs. $NP$ conjecture,
based on existence of trivial obstructions,  a {\em theoretically
infeasible strategy}--it is really just a restatement of the original problem.
Hence, the terminology  IOH (Infeasible Obstruction Hypothesis)--we also
say that  IOH  is {\em theoretically infeasible}.

Any proof strategy for  the  $P$ vs. $NP$ problem has to 
answer the following question: 

\begin{question} {\bf [The folklore version of the fundamental 
question of complexity theory]}
\label{qfeasible} 
Why should the
proof strategy  be  even  {\em theoretically  feasible}, i.e., 
lead to a proof even theoretically? 
In other words, in what sense is it fundamentally 
different from the  trivial, infeasible strategy  above
and not just a restatement of the original problem? 

In general, what is a {\em theoretically feasible proof strategy} for the 
$P$ vs. $NP$ problem? 
Does such a strategy even exist?
\end{question}

This question is  natural and not new, 
because,  in view of the universality 
of the $P \not =  NP$ conjecture, it has always been questioned right from
the beginning  why it
should even be provable. For example,  why can it not be 
independent of the formal axioms of mathematics? This was studied 
in some early works; cf. \cite{aaronsonformal} for their  survey. It 
was also the 
motivation behind identifying  the fundamental local
barriers towards this problem stated above.
These local  barriers   roughly say that the proof techniques akin to
the ones used for proving three types of lower bounds, 
namely the classical hierarchy theorems, lower bounds for 
constant depth circuits \cite{sipser} and the $P\not = NC$ result 
without bit operations [GCTlocal], would not work for the $P$ vs. $NP$ problem
in a sense because the problem  itself  stands in their way,
along with its variants and properties, 
rendering them theoretically infeasible, so to say.
Admittedly,  since these local barriers apply to 
proof techniques that constitute  only    very specific 
subareas of mathematics, 
we cannot conclude anything about the proof techniques
which go through other areas of mathematics on the basis of such   local
barriers.
But one can still wonder if the $P$ vs. $NP$ problem could not
stand in the way of every proof technique like this, rendering it
theoretically infeasible, and thereby precluding its own proof. 
That is, what happened on the earlier three occasions, why can it not
keep on  happening ad infinitum? 
This does not seem inconceivable 
since the problem itself says that the discovery
is hard, and so, why could it not make the discovery of its own proof hard?
Hence arises the fundamental folklore question
that has been lurking in the background of the $P$ vs.
$NP$ problem right from the beginning:
why should a given proof technique 
be even {\em theoretically feasible}  as per some 
reasonable notion of theoretical feasibility.

By the {\em  complexity barrier} we mean the task of answering this
Question~\ref{qfeasible} {\em formally}. The key phrase here is formally.
Because Question~\ref{qfeasible} itself is an informal folklore question
with no theorem or precise mathematics behind it. To answer it formally
means:

\begin{enumerate} 
\item[1] {\bf Formalize the question:} 
Fix a proof strategy and  formalize the question first by defining 
formally {\em theoretically feasibility} 
of this  proof strategy,
justify why this  is a reasonable formalization, and then
\item[2] {\bf Answer it:}
Answer the formalized question, i.e., prove that
the proof strategy under consideration is theoretically feasible
as per this formal definition.
\end{enumerate} 
Here step 1  may be possible 
for only  exceptional proof strategies. That is, the $P$ vs. $NP$ problem
seems so severe that it may not even allow most proof 
strategies to define formally their own theoretical feasibility, let alone
prove themselves to be theoretically feasible and come near the problem.
Thus formalization of the folklore question  as in step  1
is itself a part of the challenge, a part of the root difficulty in 
approaching the $P$ vs. $NP$ problem.

This is not a new phenomenon.
Indeed, many fundamental questions in science are informal folklore 
questions and a part or sometimes the whole of the
challenge lies in just formalizing them.
For example:

\begin{question}{\bf [The folklore version of the fundamental 
question of computability theory] }

\label{qcomputable}
What is computable (decidable)? 
\end{question} 

This is informal. 
But the main challenge in computability theory as we know now--we shall
call this the {\em computability barrier}--was to  
formalize this question by giving a precise mathematical meaning to
the phrase {\em computable} (Computability Hypothesis proposed by Church and
Turing in  \cite{church,turing})
and then justify 
that  it is a reasonable formalization. 
Of course, formalization
of the phrase computable 
that we have today is  one reasonable  formalization--we shall also call it
a {\em formalization of the computability barrier} in the form of
a computability theory. Or, we will say that 
the present computability theory  is 
one way of formalizing  the computability barrier or one formal approach
to cross the computability barrier.
We cannot say that it is the only one. There may be other ways of 
reasonably formalizing  Question~\ref{qcomputable}.
Every such way will correspond to
a different computability theory.
But  there is just one computability theory today that has been universally 
accepted as reasonable.

The formalization of this computability barrier is the key ingradient 
in proving  the fundamental Incompleteness Theorem of logic  proved by 
G\"odel  \cite{godel}--namely, that 
the language of true statements in number theory is undecidable.
Actually, the Incompleteness Theorem
predated computability theory and 
was originally  stated in terms of definability rather than computability.
But let us imagine asking
if  number theory is 
decidable, without even having a formal definition of  decidable.
(In fact, this is essentially  what happenned  in Entscheidungsproblem
posed by Hilbert \cite{hilbert} which
asked  if number theory  is decidable mechanically--and this
was just taken as a folklore notion.)
Then the fundamental difficulty in answering this question is 
just formalizing the informal Question~\ref{qcomputable}.
That is, formalization of the computability barrier above. Once this is done 
and  the phrase 
computable or decidable is given a formal meaning, it becomes possible to
prove the Incompleteness Theorem by encoding computations.
The Incompleteness Theorem 
is also a universal statement about mathematics like the $P$ vs. $NP$ problem.
Hence, it is especially relevant to study 
the  nature of the fundamental conceptual difficulty in its proof. 
As we see in retrospect,  this was basically 
formalizing the  fundamental  folklore Question~\ref{qcomputable}.
If so, it should not be surprising if at least a part of the 
fundamental conceptual difficulty in  the proof of the other 
universal statement about mathematics--the $P$ vs. $NP$ problem--should 
also be just formalizing some other  fundamental folklore  question.

Question~\ref{qfeasible} seems to be that  question.
As per the folklore, it 
is supposed to be the  root 
difficulty of the $P$ vs. $NP$ problem, and a part of this
difficulty is just formally stating what the question is saying, i.e,
formalizing this question by giving a precise 
mathematical meaning to  {\em theoretically feasibility} of a judiciously 
chosen proof strategy. 
If by complexity theory (of lower bounds) we mean a theory for approaching
the $P$ vs. $NP$ and related  problems, then there are as many complexity 
theories as there are ways of formalizing this question reasonably, because
each such formalization  chalks out a way of crossing the
complexity barrier. GCT is one such way.
But there is one fundamental difference between computability theory
and complexity theory. We do not really expect many computability 
theories--i.e. ways of crossing the computability barrier--whereas we do expect
many complexity theories--i.e.  ways of crossing the complexity barrier.

Now we have to explain what we mean 
by a reasonable  formalization of  Question~\ref{qfeasible}. We mean that 
it should be in the spirit of the reasonable formalization of 
Question~\ref{qcomputable} in \cite{church,turing}. Thus it should be 
absolutely precise and formal as in these articles, and the interpretation
of the phrase {\em theoretically feasible} should  be 
reasonable like the interpretation of the phrase {\em computable}
as in these articles. Specifically, we mean:

\begin{enumerate} 
\item[I] Give a mathematically precise formulation of  an  Obstruction
Hypothesis (OH) that is meant to be an alternative to IOH.
Here it must be proved unconditionally that OH implies  IOH.
Furthermore, this obstruction hypothesis should be of the form: 
\[({\bf OH}): \quad  \forall n \forall  m=\poly(n): s(n,m),\] 
where $s(n,m)$ is some formal mathematical statement. 
The statement $s(n,m)$ is supposed to  say that when $m=\poly(n)$ there 
exists  some cause of hardness (which we shall call an obstruction) 
because of which IOH holds. We put no restriction on the nature 
of this cause. The cause could be existence of some
mathematical object in the spirit of a trivial obstruction, in which
case this object can be called an obstruction--an alternative to a
trivial obstruction.
But this is not necessary. The cause could be just some 
mathematical condition.
As long as it can be completely formalized in the form of a precise 
mathematical statement $s(m,n)$ and OH implies IOH unconditionally,
we do not care. We call OH a
{\em complexity theoretic hypothesis} because it is necessary that
$m$ be small in comparison to $n$ for it to hold.

\item[II] 
State precise formal {\em mathematical hypotheses} 
MH1, MH2, ... , MHr, $r \ge 1$, and 
give a  mathematically  precise  formal  and reasonable definition of the
phrase {\em theoretical feasibility} (of OH)
such that if MH1,...,MHr hold then OH becomes theoretically 
feasible as per this definition.
Here theoretical feasibility of OH 
is supposed to mean that once MH1, ..., MHr
are proved  the $P\not = NP$ 
conjecture is not expected to stand in the way of proving OH thereafter.
Each mathematical hypothesis should be of the form 
\[ {\bf (MHi):}  \forall n \forall m t_i(m,n), \] 
for some formal mathematical statements $t_i(m,n)$.
We are calling it mathematical because this is supposed to hold for
all $n$ and $m$, not just when $m=\poly(n)$ or small as in the
case of the complexity theoretic IOH or OH. 
In other words,  smallness  of $m$ with respect to  $n$, the 
complexity theoretic crux of the lower bound  problem under consideration, 
is supposed to be absent  in the mathematical hypotheses.

\item [III]
(a)  Give reasonable justification for  why OH should hold assuming IOH
(which we take on faith as true) or some reasonable stronger forms of IOH.
This is certainly a circular 
argument. Its only function here is to give us reasonable confidence
that  OH is actually true, i.e., there is no loss of truth in going 
from IOH to OH.

(b)  Give reasonable justification for why MH1,...,MHr should 
hold.

(c)  Give reasonable justification for why 
the $P \not = NP$ conjecture or some equally hard or harder 
complexity theoretic hypotheses  should
not stand in the way of proving MH1,..., MHr and
in the way of proving  OH once MH1,..., MHk are proved.
In other words, all these hypotheses should only involve ``easy'' 
complexity theoretic hypotheses, where by ``easy'' we mean the $P\not = NP$ 
conjecture or its equivalent or harder variants 
are not expected to stand in the way of their proofs.

At least one intuitive 
reason for why the $P\not = NP$ conjecture should not
stand in the way of proving the  mathematical hypotheses is already there:
namely, smallness of $m$ with respect $n$, the complexity theoretic crux of 
this  conjecture, is absent in these hypotheses. But just this is not enough. 
It is also necessary to give additional justifications as to why some
hard complexity theoretic hypotheses are not hidden or encoded within these 
mathematical hypotheses or  OH in some hard to recognize forms. 

\end{enumerate} 

\noindent {\em Remark:} In steps I,II, and III above 
we can substitute the usual uniform version of the $P\not = NP$ 
conjecture in place of IOH. That is, we do not really require here that 
the approach has  try to prove the stronger 
nonuniform version of the $P\not = NP$ conjecture. The usual uniform
version will do.

Once steps I to III are carried out  the 
``hard'' (i.e. theoretically infeasible) complexity
theoretic hypothesis IOH is transformed into 
complexity theoretic hypothesis OH which becomes 
``easy'' (i.e. theoretically feasible)  once 
the mathematical hypotheses MH1,...,MHk are proved, and these 
mathematical hypotheses too involve only ``easy'' complexity 
theoretic hypotheses, such
that the $P \not = NP$ conjecture is not expected to stand in the way of
the proofs of any of these ``easy'' complexity theoretic hypotheses.
Thus once Question~\ref{qfeasible} is fully formalized in this fashion,  hard 
complexity theory (IOH) is exchanged with hard mathematics 
(MH1,...,MHr + OH). What is gained in this exchange is that 
the $P \not = NP$ conjecture is not expected to stand in the way 
doing this hard mathematics, and thus a fully formalized  mathematical 
program towards proving this  conjecture is chalked out.
We say that the {\em complexity barrier is formalized} (in this approach)
once steps I to III are carried out and thus the folklore
Question~\ref{qfeasible} is formalized.

We  say that {\em the complexity barrier is crossed} (in this approach)
and that OH is {\em theoretically feasible} once MH1,...,MHr
are proved--we also say that the proof strategy (approach) becomes 
theoretically feasible. Theoretical feasibility of the proof strategy 
does not automatically guarantee 
{\em practical feasibility}  of proving OH. That is, even after OH has
been proved to be theoretically feasible actually proving it may be
a mathematical challenge, though  it is a formally ``easy'' complexity
theoretic hypothesis then.
This is why we regarded  OH  as hard mathematics above.

Now arises the  next  question: 
what is meant by reasonable justifications in III? 
Certainly we cannot expect a proof technique to formally 
prove that the $P \not = NP$ 
conjecture or its harder variants 
would not stand in the way of proving MH1,...,MHr and
$OH$, because the only way to prove this formally is to actually prove them 
unconditionally thereby proving the $P \not = NP$ conjecture itself.
But this is a serious issue because regardless of how reasonable 
the justifications of a proof technique may look at the surface, 
what is the guarantee that 
the $P\not = NP$ conjecture or some 
equivalent  or harder complexity theoretic assumptions, such as 
existence of one way functions, pseudorandom generators etc. as per
the hardness vs. randomness principle \cite{nisan}, are not 
lying hidden within MH1,...,MHr and OH 
in some hard to recognize forms?  That is, what is the 
guarantee   that
the proof technique is not simply restating 
the problem in an equivalent terms  with  an ingenious encoding
and thus going   in circles, or worse,  making it harder
and thus going backwards, thereby simply evading 
Question~\ref{qfeasible} despite all this formalization?

To see the seriousness of this issue, let us imagine ourselves  in the 
early seventies, before the work on pseudo-random generators, derandomization,
one-way functions and so forth began, and imagine that  some proof technique 
came along claiming to be an approach to the $NEXP$ vs. $P/poly$ 
problem--which is far, far weaker than the $P$ vs. $NP$ problem,
but let us ignore that--by  reducing this problem to 
the problem of showing that polynomial 
identity testing belongs to $P$ and then claiming to be theoretically
feasible because the latter is an upper bound problem involving $P$ which
stands for theoretically feasible. At that time  this
may have seemed like a reasonable justification. After all, if a proof
technique has reduced a lower bound problem to an upper bound problem 
over $P$, then it seems like  progress, at least, at the surface. 
Coming to back to the present, we know now, thanks to all the work on
the hardness vs. randomness principle \cite{hastad2,nisan,russell}
that has happenned in the meanwhile,
that the proof technique was just going in circles. Because polynomial
identity testing belongs to $BPP$ and as per \cite{russell} derandomization
of polynomial identity testing essentially amounts to proving the
original lower bound problem or worse the  harder permanent vs. determinant
problem. In other words, the hardness vs. randomness principle tells us today
that all arguments which look like reductions to 
upper bound problems at the surface 
but involve pseudorandom generators, one way functions, derandomizations,
and so forth  are circular, 
something that we may not have seen so clearly in the seventies. 

Indeed, one significant
outcome of the works on $NP$-completeness \cite{cook,karp,levin},
plausible formal independence of the $P$ vs. $NP$ problem 
(surveyed in \cite{aaronsonformal}),  the  local
barriers \cite{solovay,rudich,GCTlocal,aaronson},
and the hardness vs. randomness principle 
\cite{hastad2,nisan,russell} in the last few decades 
was the discovery of the formidable  circle around the $P$ vs. $NP$ 
problem. That is, they together raised Question~\ref{qfeasible} and 
pointed  out the extreme difficulty of penetrating this circle.

Which is why, though we cannot ask 
for a complete guarantee that the $P\not = NP$ conjecture or its
variants would not stand in the way of proving MH1,...MHr, and OH,
because that only comes
after proving $P \not = NP$,  we need at least some reasonably {\em hard}
guarantee that the proof technique has indeed penetrated 
this  circle.

To see how to get such a hard guarantee, let us closely compare 
the circle around the $P$ vs. $NP$ problem with  the one around 
the Incompleteness Theorem and then try to
transfer what we already know about how the circle around the latter was 
penetrated to get an idea about how the proof technique 
that penetrates the circle around the former ought to be.
There is a circle around the Incompleteness Theorem 
because when it says there exist unprovable statements
in mathematics or that the truth in number theory is undecidable 
it {\em seems} as if it is also hinting that it may be
unprovable itself or that its truth may be undecidable. There is a circle 
around the $P$ vs. $NP$ problem, because when it says that 
discovery is hard it {\em seems} as if it is also hinting that the discovery
of its own proof may be hard. Any proof strategy for the Incompleteness
Theorem or the $P$ vs. $NP$ problem has to break this
self-referential paradox--the circle. The conceptual breakthrough in 
\cite{godel} was to break this circle for the Incompleteness 
Theorem and show that it is talking about the rest of mathematics not itself.
If we closely examine the proof strategy in \cite{godel}, we see that it
is {\em extremely rigid}.
By this we mean that we know just one way of penetrating this circle
today, and though there are many versions and proofs of the Incompleteness
theorem, all of them essentially use the same basic idea.
Similarly, the formalization
of the the computability barrier  in \cite{church,turing}--the key 
ingradient in proving the Incompleteness Theorem--is also 
{\em extremely rigid}.
By this, we mean
we know just one reasonable formalization of the computability barrier today.
So far, we do not know any other formalization 
that is ``more or less reasonable'', ``close to reasonable'' or 
``almost reasonable''. 
In short, it is a $0$-$1$ game: it either works completely or fails completely.
We expect this $0$-$1$ game to continue 
in the $P$ vs. $NP$ problem as well.

This $0$-$1$ game cannot be  formalized.
But we can  state a certain   Rigidity Hypothesis formally which 
intuitively suggests this $0$-$1$ game for the complexity barrier as well.
Towards that end, let us assume that the proof technique has 
provided  as in the
the natural proof paper \cite{rudich} 
a formal statement 
of a Useful Property ($UP_n$), which lies at its heart, and assuming which 
it plans to prove $P \not = NP$. Formally, $UP_n$ is a subset of the set
of all $n$-ary boolean functions. Its definition is 
as in \cite{rudich} and we do not restate it here. 
Let $N=2^n$ be the truth table size of specifying any $n$-ary boolean 
function.

\begin{defn} \label{drigidcomplexity}
We say that a proof technique 
is 
\begin{enumerate} 
\item 
{\em Nonrigid} if  $|UP_n| \ge 2^N/N^c$, for some constant $c>0$,
\item {\em Mildly rigid} if $|UP_n| \le 2^N /N^c$ for every constant $c>0$,
as $n\rightarrow \infty$,  or
more strongly, $|UP_n| \le 2^N/2^{n^a}$, for some constant $a\ge 1$,
\item {\em Rigid} if  $ |UP_n| \le 2^{\epsilon N}$,
for some $0 \le \epsilon < 1$, 
\item {\em Strongly rigid} if 
$|UP_n| \le 2^{N^\epsilon}$, for some $0 \le \epsilon < 1$,  and 
\item {\em Extremely rigid} if 
$|UP_n| \le 2^{\poly(n)}$, i.e., $\le 2^{n^a}$ for some constant $a>0$.
\end{enumerate}
\end{defn} 

Thus probabilistic proof techniques to which the natural proof
barrier \cite{rudich} applies are nonrigid, and 
if a proof technique is mildly rigid it bypasses this
barrier, since it violates the largeness criterion in \cite{rudich}. 

\begin{hypo}[Rigidity Hypothesis (RH)] 
Any eventual proof (technique) that shows $P\not = NP$ is 
extremely rigid.
\end{hypo} 

The natural proof paper says that any eventual proof of the $P\not = NP$ 
conjecture is mildly rigid (assuming 
some hardness assumptions and an additional constructivity criterion). 
The Rigidity Hypothesis says  that the reality is extremely 
severe compared to this. Its detailed discussion with motivations etc.
will appear in GCTflip.

Computability Hypothesis \cite{church,turing} says that 
there is essentially just one reasonable way to formalize the computability 
barrier and that way--i.e., the specific notion of computability that 
we have today--is extremely rigid as explained above. 
In the case of the complexity barrier there need not be just one 
way to formalize it. The Rigidity Hypothesis suggests that that 
each way may be  extremely rigid again--i.e., the notion 
``a theoretically feasible proof technique''
may  also be an extremely rigid notion like ``computable''.
Unfortunately, we cannot formally define extreme rigidity of a notion. 
The Rigidity Hypothesis is the best that we can do formally. 
It is talking about extreme rigidity of a proof, but that may be taken
as suggesting extreme rigidity of the notion of 
``a theoretically feasible proof technique'', i.e., extreme rigidity of 
formalizing  the complexity barrier 
as in the case of the computability barrier.
Intuitively, this suggests that 
a formalization of the complexity barrier may be  either 
completely right or completely wrong like in the case of the computability
barrier. 
Which is why if a formalization  is reasonable  and
extremely rigid, then intuitively we 
expect only  two possibilities: either 
it is really an approach to the $P$ vs. $NP$ problem or just 
a circular  restatement, and we have to distinguish between the two. 
For this, all we have to do is to insist that some restricted form 
of the proof strategy  also 
yield an  unconditional  lower bound that is nontrivial  enough so that if
the proof strategy    were circular it will simply not be able to 
provide such a lower bound, and furthermore, we also insist that 
the proof strategy  should  provide evidence that 
it  bypasses the earlier known local barriers.
If it does this, then assuming that it is a  $0$-$1$ 
game, there is a reasonable assurance that 
the  formalization is indeed an approach to the $P$ vs. $NP$ problem.

With this  in mind, 
we will say  that a proof strategy  is a {\em formal
approach towards the $P$ vs. $NP$ problem} if it meets the
following criteria: 

\begin{enumerate} 
\item[A]  Some weaker form of the  proof strategy yields 
a concrete unconditional superpolynomial 
lower bound that  is a 
restricted form and a formal implication 
of any of the fundamental separation problems 
in Figure~\ref{fig:complexityclasses} 
(the center of complexity theory)
in a realistic and natural model of computation, where by natural and
realistic we mean a model that is actually used extensively in practice and is
not defined for the sake of proving  the lower bound.

\item[B] The proof strategy provides evidence that it 
bypasses the earlier known  
local barriers mentioned before simultaneously.
Here, in view of the fundamental nature of these barriers and given
that they were proposed and accepted in the field as metabarriers,
we agree to take them in the meta sense--i.e., in  spirit rather than letter. 
Because formally  the natural proof barrier \cite{rudich}  is
defined only for the unrestricted $P$ vs. $NP$ problem and
is crossed only after proving $P\not = NP$, which is of no help 
in evaluating any proof strategy. Fortunately, this and other local barriers
make sense in spirit in  any reasonable restricted setting. 
Thus we will say that 
a proof technique bypasses the natural proof barrier \cite{rudich} 
if it does not
apply to most functions, with the meaning of most like in \cite{rudich}
(this makes sense over the boolean field as well as over $\Q$ as long as
we take bitlengths of rational numbers into account in a natural way, and 
if a restricted model is natural enough--which we require--it 
should come with  a natural meaning for most).
We will say that it bypasses the relativization barrier \cite{solovay} 
if it is algebraic, 
i.e., works on algebrized forms of computations  in the spirit 
of the $IP=PSPACE$ result.
We will say that it bypasses the algebraic degree barrier [GCTlocal] if it 
does not use low degree techniques.

\item[C] It provides a reasonable formalization of Question~\ref{qfeasible}
by carrying out steps I,II and III above.

\ignore{We also require that the  OH in the proof of the lower bound in B be 
a natural restricted form of the OH in this formalization.}

\item[D] (Extreme Rigidity): It formally states 
the UP for the $P$ vs. $NP$ problem and proves its extreme rigidity.
It should be clear that this UP 
lies at the heart of  the approach. 
\end{enumerate}

We shall refer to these as the criteria A,B, C, and D to determine if
a proof strategy is a formal approach.

All approaches can agree on A and B given that the fundamental 
difficulty of proving any natural and realistic superpolynomial lower bound
implication of any of the fundamental separation 
problems in  Figure~\ref{fig:complexityclasses} 
and of crossing the earlier local barriers simultaneously has been
well recognized by now. An approach which is just restating the $P$ 
vs. $NP$  or related problem, such as the hypothetical proof strategy 
in the seventies above, by hiding the original lower bound problem
or its equivalent or harder 
forms in any of its hypotheses will not be able to meet criteria A and B.
Hence if an approach can meet the criteria
A and B, this provides a 
reasonable hard guarantee  that it is not restating the original problem  and 
going in circles, which is  what crossing the complexity  barrier
finally means: overcoming the  seemingly self-referential nature of the
$P$ vs. $NP$ problem. 
Similarly, all approaches can agree on C given
that it has been accepted in the folklore that Question~\ref{qfeasible}
is the root difficulty of the $P$ vs. $NP$ problem and that any 
approach would have to finally answer it in one way or the other. 
But if a proof technique  cannot  even formally 
state the question that it is trying to answer how can we call it an approach? 
Motivation behind D is the Rigidity Hypothesis stated above.

Once A, B, C, D are met, the proof technique 
has chalked  out a complete 
mathematical program C towards the $P$ vs. $NP$
problem in the form of the hypotheses MH1,...,MHr and OH, provided 
reasonable  justification  that the $P\not = NP$ conjecture should not 
stand in the way of proving MH1,...,MHr and hence OH,  and 
has provided  a reasonable hard guarantee  in the form of the criteria
A and  B  that this path is actually going
forwards,  not in circles, or even worse, backwards, and has shown
that the program is extremely rigid. It is then
reasonable to call it  a formal approach to the $P$ vs. $NP$ problem
in view of the Rigidity Hypothesis and the  remarks  before and
after that hypothesis above.

By a {\em complexity theory of the $P$ vs. $NP$ problem} we mean 
a mathematical theory of any formal approach towards this problem. 
Just as we have as many computability theories as there 
are formal approaches 
to cross the computability barrier, there are as many complexity 
theories of the $P$ vs. $NP$ problem as there are formal approaches to
cross the complexity barrier.
But unlike in the case of the computability theory, we expect
many complexity theories--though as we have already said, 
the $P\not = NP$ conjecture seems  so severe that each such complexity
theory may have to be exceptional (like the present computability theory) 
because otherwise formalization of the complexity barrier may simply be
impossible (as in the case of the computability barrier).

The complexity barrier can also be defined for other problems related 
to the $P$ vs. $NP$ problem, such as the $P$ vs. $NC$ or $\#P$ vs. $NC$ 
problems. Formal approaches and complexity theories associated with
these problems are defined similarly.

It may be noticed that the meaning of the phrase barrier in
complexity barrier is different from its meaning in the local barriers
such as the algebraic degree barrier. The algebraic degree barrier has a 
a precise and formal mathematical  definition, cf. Chapter 7 in GCTlocal.
In contrast, formalizing the complexity barrier is in fact 
a part  of that  barrier.
The complexity barrier is a 
barrier in the same sense that the computability barrier is a barrier.
There is no computability theory until the computability barrier is
formalized. Similarly, if we accept that Question~\ref{qfeasible} is the 
root difficulty of the $P$ vs. $NP$ problem, then there is no 
complexity theory of the $P$ vs. $NP$ problem in a formal sense 
until the complexity barrier is formalized.
In that sense we may say that the complexity barrier is the main
barrier of complexity theory. Whereas the algebraic degree barrier 
is more like a guiding post.
As long as we keep in  mind that  universal and 
local barriers are of a different kind, there should be no confusion.

\section{The main result of GCT} \label{sGCT}
The main result of GCT is that it is a formal approach towards the $P$ vs. 
$NP$ and related problems that meets the criteria A,B, C,  and D. This then
yields one formal complexity theory of the $P$ vs. $NP$ problem.
The $P\not = NC$ result without bit operations [GCTlocal] meets 
criterion A as explained in Section~\ref{swithoutbit}. 
The mathematical form of the $\#P\not = NC$ conjecture in characteristic 
zero [GCT1,2,6]  meets criterion B as explained in Section~\ref{smathform}. 
The main result of GCT6 gives a reasonable formalization of the 
complexity barrier; i.e., reasonable formalization of  
Question~\ref{qfeasible}; cf. Sections~\ref{sgct6}-\ref{sexpli}.
This meets criterion C.  Criterion D for the permanent vs. determinant 
problem is addressed  in
Section~\ref{scharsym} and for the $P$ vs. $NP$ problem in  GCTflip.

We have already remarked that the complexity barrier plays the same role
in the context of the $P$ vs. $NP$ problem that the computability 
barrier plays in the context of the 
Incompleteness Theorem.  GCT6 formalizes  the 
complexity barrier in the same spirit 
that the articles \cite{church,turing} formalize  the
computability barrier. 
Hence it is illuminating to compare the 
formalization of the computability barrier 
in \cite{church,turing}  with 
the  formalization of the complexity barrier in GCT6. 
The formalization of Question~\ref{qcomputable} 
in \cite{church,turing} is mathematically elementary
(though conceptually a breakthrough) whereas the formalization of 
Question~\ref{qfeasible}  in GCT6 is mathematically nonelementary, because 
it crucially depends on fundamental results and constructions in 
algebraic geometry and representation theory. At present, we do not
know any elementary  way of formalizing Question~\ref{qfeasible} reasonably.
Once Question~\ref{qcomputable} is formalized, the proof of the Incompleteness
theorem in \cite{godel} (or rather its translation by substituting
computability for definability)
is also mathematically  elementary  (though conceptually a 
breakthrough again).
In contrast, the plan for answering the formalization of 
Question~\ref{qfeasible}
in [GCT6,7,8,11] goes through the theory of nonstandard quantum groups 
and seems to need nontrivial extension of the work surrounding 
the profound positivity result of mathematics--the Riemann hypothesis
over finite fields \cite{weil2}. 

This suggests that there is a massive gap in the  difficulty of the
Incompleteness Theorem and the $P$ vs. $NP$ problem. That there should be
a gap  should not
be surprising. The former is a universal statement about  the idealized 
world of mathematics wherein the computational complexity measures--time
and space--do not matter, whereas the latter is a universal statement 
about the far more complex physical world of mathematics wherein time and
space do matter. But what may be surprising perhaps is the massive scale of
the gap. 

One may ask if this massive gap  between the two 
universal statements about mathematics is intrinsic, or if there is an easier 
elementary formal approach to the $P$ vs. $NP$ problem wherein the gap
is not so huge.
Well, at present we are not aware of any 
unconditional lower bounds in complexity theory 
that meet the criteria A and/or  B other than the ones in GCT mentioned above.
The proof of the first lower bound of GCT 
needs classical algebraic geometry, whereas 
that of the second geometric invariant theory\cite{mumford}. 
Furthermore, as we mentioned above, the formalization in GCT6 
of Question~\ref{qfeasible}--the root difficulty of the $P$ vs. $NP$
problem-- also needs algebraic geometry and representation theory. 
That is, at present, 
we cannot even state formally in an elementary language what 
Question~\ref{qfeasible} is saying let alone set up some elementary 
program for answering it. Furthermore, universality means that 
the basic structure of all the  approaches towards the $P$ vs. $NP$ problem
should be the same, just as the basic structure of  all the proofs of 
the Incompleteness Theorem and related results known so far is the same.
Thus it seems that the $P$ vs. $NP$ problem may be
a  difficult problem after all  that needs  nonelementary 
techniques for  any realistic chance of progress 
despite its  deceptively elementary  statement.

The rest of this article gives  an overview of how GCT meets the criteria
A, B, C,  and D.  The formalization of the complexity barrier is done in steps.
In Section~\ref{snonuniform}, we define  abstract nonuniform approaches
towards the $P$ vs. $NP$ problem. 
Here abstract means not formal. So an abstract approach is a high level
scheme for an approach. In Section~\ref{sflip}, we restrict 
an abstract  nonuniform
approach to what   we call an (abstract) flip, which 
is the  defining abstract strategy of GCT. 
A local form of the abstract flip is implemented  in 
a restricted setting in the form of the  first lower 
bound in 
Section~\ref{swithoutbit}. The abstract flip is implemented  in a restricted
setting in the form of  the second lower bound in 
Section~\ref{smathform}. Sections~\ref{sgct6} and \ref{sexpli} 
give a high level  overview of a formalization of the  abstract 
flip (formal flip) in the unrestricted 
setting, i.e., formalization of Question~\ref{qfeasible}. A full
description of this formalization (in characteristic zero) can be
found in the mathematical overview \cite{GCTriemann}; for the 
full formalization (for  fields of positive characteristic, including
the finite boolean field as in the usual case) see GCT6.

\section{Nonuniform approaches} \label{snonuniform} 
In the rest of this paper we confine our attention to
nonuniform approaches unlike the uniform approaches used to prove the 
classical hierarchy theorems, 
since the relativization barrier \cite{solovay} 
suggests such  approaches 
have a fundamental limitation in the context of the $P$ vs. $NP$ problem.
If there exists a nonrelativizable uniform approach, we
leave the task of formalizing the complexity barrier to that approach. 
In this section we define an abstract nonuniform approach to the nonuniform
$P$ vs. $NP$ problem  in
Section~\ref{scomplexity}. Recall that the  problem is to  show
nonexistence of small circuits  for computing $f(X)=f(x_1,\ldots,x_n)$
 as there.

By  a nonuniform approach we mean 
it shows existence of some obstruction $O_{n,m}$ 
(proof-certificate of hardness as per that approach in the spirit of
the trivial obstruction  but better) depending on 
$n$ for every $n$ and $m=\poly(n)$.
All proofs of nonuniform lower bounds so far,
e.g. \cite{sipser,monotone,GCTlocal}, prove existence of such obstructions
that are better than the trivial obstructions in the lower  bound problems
under consideration.
For example, in the proof of the lower bound for constant depth 
circuits   an obstruction is a  table that lists 
for every constant depth small size circuit 
a counter example that could  be potentially produced by the derandomized 
polynomial time  version \cite{agarwal} of the random restriction procedure 
\cite{sipser}.
This obstruction  is better than the brute force  trivial obstruction in the
sense that its each row can be constructed in polynomial time instead 
of  exponential time as in   the brute force method. 

Formally:

\begin{defn} \label{dobst}
We say that a family ${\cal O}=\cup_{n,m} {\cal O}_{n,m}$ of mathematical
objects is an {\em obstruction family} to efficient computation of 
$f(X)=f(x_1,\ldots,x_n)$ if
\begin{enumerate} 
\item Each object $O$ in ${\cal O}$ called an {\em obstruction} has 
a specification (a bit string)  denoted by $[O]$. 
An object $O_{n,m} \in {\cal O}_{n,m}$ is called an obstruction to
computation of $f(X)$ by circuits of size $m$.
\item  Nonemptyness of ${\cal O}_{n,m}$, i.e.,
existence of an obstruction $O_{n,m}$ serves as a
guarantee that $f(X)$ cannot be computed by circuits of size $m$.
Specifically, there is a decidable decoding algorithm that given $n,m$ 
and the specification $[O]=[O_{n,m}]$
of an obstruction  $O=O_{n,m}  \in {\cal O}_{n,m}$ 
produces a trivial obstruction to computation of $f(X)$ by circuits 
of size $m$, i.e., a table that gives for each circuit $C$ of size 
$m$ a counter example $X$ so that $f_C(X)\not = f(X)$. 
This decoding algorithm must be guaranteed to be correct. 
That is, it must produce such a trivial obstruction correctly 
for every obstruction. But we put no restriction on the computational
complexity of the algorithm.
\end{enumerate}
Given an obstruction $O \in {\cal O}$, we let $\bitlength{O}$ denote
the bitlength of its description $[O]$.
We say that an obstruction family is uniform if $\bitlength{O}$ is $O(1)$
for every obstruction $O$.
\end{defn}

In principle, we do not need a decoding algorithm here. All we 
need is a guarantee than nonemptyness of ${\cal O}_{n,m}$ implies
that $f(X)$ cannot be computed by circuits of size $m$. But decoding 
algorithm makes everything concrete, so we will assume that it is there.

The most important  obstruction is a final proof of the $P \not = NP$ 
conjecture. It has  $O(1)$ length and can be verified in $O(1)$ time.
We call it the final obstruction.
For this obstruction there is a {\em trivial
decoding algorithm}, namely, for each circuit $C$ of size $m=\poly(n)$,
just compute $f_C$ and $f$ on all $X$ of length $n$ in any order and produce 
the first  $X$ on which they differ. A final proof of the $P \not = NP$ 
conjecture guarantees that this algorithm is correct, but it is not until then.
Once there is a final proof, an $O(1)$-size program for $f(X)$,
or for that matter any NP-complete function,
also becomes an obstruction with a trivial decoding algorithm,
but not until then.
Other well known $O(1)$-size obstructions in the $P$ vs. $NP$ 
problem are polynomial time  one way functions 
or  pseudo random generators.  But the problem of proving existence 
of such obstructions is  believed to be 
even harder than the $P$ vs. $NP$ problem.
Alternatively, we can consider an $O(1)$-size program for an explicitly given
conjectural one way function or pseudo random generator. This
is  an obstruction only 
after it is proved that the  function under consideration
is indeed a one-way function or a pseudo random generator,  but 
proving that  is again believed to be   harder than the original 
$P$ vs. $NP$ problem. Thus it can be seen that 
any argument based on such  uniform, i.e., $O(1)$-size objects is bound to be
circular, which is what has to be avoided to cross the complexity barrier,
and furthermore,
the relativization barrier \cite{solovay} 
also suggests that arguments based on uniform
objects may have fundamental limitations.
Hence, our  interest   is primarily in the
intermediate obstructions that are meant to be a vehicle in 
a given nonuniform approach  to reach 
a final proof (i.e. a final obstruction)
in the spirit of the intermediate  obstructions that are 
used to prove the earlier nonuniform lower bound results
\cite{sipser,monotone,GCTlocal}. Hence,
whenever we say obstructions  we have in mind intermediate 
nonuniform obstructions
which lie on the way from the trivial to the final obstruction.

\section{The flip: going for explicit construction} \label{sflip}
We now define the abstract defining strategy of GCT, called the flip 
[GCT6,GCTflip],
for formalizing and crossing the complexity barrier.
It is the most obvious and natural nonuniform strategy for this
suggested by the $P$ vs. $NP$ problem itself. 

\begin{figure} 
\begin{center}
\epsfig{file=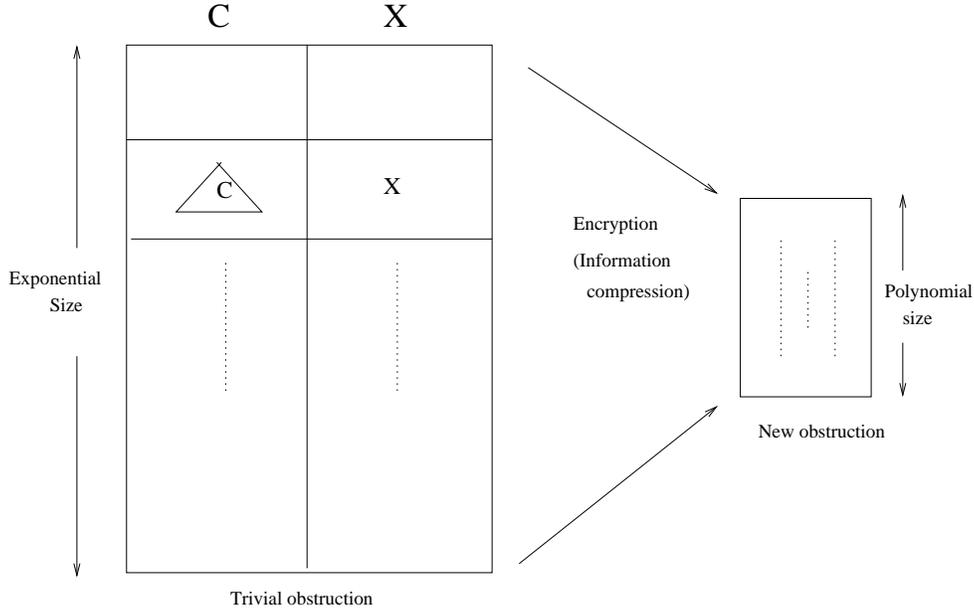, scale=.6}
\end{center}
      \caption{From trivial to new obstruction}
      \label{fig:trivialtonew}
\end{figure}

Before we define it, 
let us first see what is wrong with the trivial obstruction from the 
complexity-theoretic point of view (cf. Figure~\ref{fig:trivialtonew}).
That is quite clear. First, it is long,
i.e.,  its description takes exponential space. Second, it is hard to
verify (and also construct); i.e., this  takes exponential time.
Since $NP$ is the class of problems with ``proof-certificates'' that 
are short of polynomial-size and easy to verify in polynomial-time, 
this then leads to the following flip strategy
for proving the nonuniform
$P\not = NP$ conjecture, with the obvious complexity-theoretic
interpretation of the phrase theoretically feasible.
(We treat formalization and crossing of the complexity barrier 
simultaneously below, instead of separately as we did in
Section~\ref{scomplexity}).

\noindent {\bf I. (The flip: theoretical feasibility of verification) :}

\noindent (a) Formulate a new notion of a (nonuniform)
obstruction (cf. Definition~\ref{dobst})
that is {\em explicit}, by which 
we mean {\em short and easy-to-verify}.
Formally:

\begin{defn} \label{defnobstexplicit}
We say that an obstruction family ${\cal O}$ (Definition~\ref{dobst}) 
is {\em explicit} if:
\begin{enumerate}
\item {\bf Shortness:}
For any $n$ and $m< 2^n$, if there exists  an 
obstruction in ${\cal O}_{n,m}$ 
to efficient computation of $f(X)$ by circuits of size $m$,
then there also exists a short  obstruction $O_{n,m}$ in ${\cal O}_{n,m}$, 
where by short we mean 
the bitlength $\bitlength{O_{n,m}}$ of its specification is
$\poly(n)$ (note the bound does not depend on $m$).
\item {\bf Easy-to-verify:} Given $n$ in unary, $m$ in binary 
and a bit string $x$, whether 
it is the specification $[O_{n,m}]$ of an obstruction  
$O_{n,m} \in {\cal O}_{n,m}$ can be decided  in $\poly(n,\bitlength{x})$
time, where $\bitlength{x}$ denotes the bitlength of $x$
(note the bound again does not depend on $m$).
\end{enumerate}
\end{defn} 

This means that the problem of verifying (recognizing)
an obstruction (or rather its specification) belongs to $P$, and hence,
is theoretically feasible (``easy''). Thus this notion 
of an obstruction is then fundamentally better than the theoretically
infeasible trivial obstruction as far as verification is concerned.

Now, to prove $P \not = NP$ it  suffices  to show that 
${\cal O}_{n,m}$ is nonempty if $m=\poly(n)$, $n \rightarrow \infty$.
In this step, the approach should also justify on the basis of
mathematical evidence why this should be
so assuming  $P \not = NP$ (nonuniform).
This circular reasoning at this stage is only meant to convince us 
before we  plunge ahead that such obstructions should exist, i.e., there 
is no loss of essential information,  
assuming $P \not = NP$, which we take on faith, and the goal
now is to prove the existence of these obstructions.

\noindent (b) Also state precise formal 
mathematical hypotheses MH1, ..., MHs assuming which 
the problem for verifying obstructions belongs to $P$. (Here the meaning
of a  mathematical hypothesis  is as in Section~\ref{scomplexity}). 
Also justify (as in the step II of Section~\ref{scomplexity})
why the $P\not = NP$ conjecture should not stand in the way of proving these
mathematical hypotheses. 

By a reasonable formalization of this step I of the flip,
we mean precise mathematical formulation of MH1, ..., MHs, with a 
reasonable justification for  why the $P\not = NP$ conjecture 
should not stand in the way of their proofs.
By  implementation of this step I, we mean 
formally proving MH1, ..., MHs and thereby proving that 
the problem of verifying obstructions belongs to $P$.

Intuitively (cf. Figure~\ref{fig:trivialtonew}), we can think of 
the new  explicit 
obstruction as an 
encryption of a some trivial obstruction of exponential
size which is highly compressible
in the information theoretic sense--i.e., such that 
its description can be encrypted in polynomial size--and furthermore,
such that the encrypted information can also be verified fast in polynomial
time.
At the surface, it is hard to believe that 
such an obstruction can even exist. To see why, let us note that 
to verify if $f(X)=f_C(X)$, we have to evaluate 
$f_C(X)$,   cannot be  done   in polynomial time, i.e., in
$n^a$ time for some fixed $a>0$,  if the size $m$ of
$C$ is superpolynomial (say $n^{\log n}$) or alternatively,
if $m=n^b$ where $b$ is much larger than $a$. In other words,
if $a$ is fixed and we want to evaluate circuits of polynomial size
of nonfixed degree we are in trouble and 
even one row of a trivial obstruction cannot be evaluated in
$n^a$ time for any fixed $a$
To verify the new obstruction,
in effect, we  are 
{\em implicitly} verifying   all exponentially many  rows of some 
trivial obstruction  in $O(n^a)$ time for some fixed $a$, but without 
even knowing what that trivial obstruction is, since the verification 
procedure cannot afford to decrypt the new obstruction using
the associated decoding procedure to get an exponential
size trivial obstruction. This may seem impossible at the surface.
In fact, this may seem impossible even 
if we were to allow ourselves $\poly(m)$ time, because verification of 
a trivial obstruction given in the form of an oracle--let us call it
$O$--is a 
$\Pi_2^O$-problem, which we do not expect to belong to $P$ if $P \not = NP$.
Thus it may seem that 
such a scheme could not possibly exist and that the $P \not = NP$ conjecture
would again stand in the way of this verification 
either directly or indirectly (implicitly).
Formalization  of  this step of the flip--i.e. formulation
of  MH1, ..., 
MHs with a reasonable justification for why the $P\not = NP$ conjecture 
should  not stand in the way of their proofs--amounts to   justifying why
the $P\not = NP$ conjecture
should  not stand in the way of verifying the new obstruction--which 
is what formalizing  the complexity barrier
essentially means here.

Of course, the actual implementation 
does not have to  encrypt any trivial obstructions. The 
new obstruction could be such that its existence implies existence of
some trivial obstruction in a highly indirect way. That is,
the computational complexity of the associated decoding algorithm 
(which we are only assuming is decidable) could be extremely high. 
And for that matter, we do not even need such a decoding algorithm 
as long as there is a guarantee that existence of an obstruction implies
existence of a trivial obstruction.
But if it is possible to get a trivial obstruction from a new obstruction
easily; i.e., if the associated {\em decoding algorithm} works in polynomial
time, by  which we mean 
given specification of 
a new obstruction $O_{n,m}$ and a small $C$ of size $m$ it
produces in $\poly(n,m, \bitlength{O_{n,m}})$ time a counterexample $X$ for 
$C$--then we say that the new obstruction is {\em easy to decode}. 
Note that if $O_{n,m}$ is short, i.e., $\bitlength{O_{n,m}}=\poly(n)$,
then this running time is $\poly(n,m)$.
Though we do not need the decoding algorithm to work in
polynomial time, at least at this stage, it may intuitively
help to imagine that the new obstructions are easy to decode.

It may be noticed that a final proof of the $P \not = NP$ conjecture 
is  an explicit obstruction for every $n$ and $m=\poly(n)$ 
as per Definition~\ref{defnobstexplicit} since it can be verified 
in $O(1)$ time. That is how it ought to be since a final proof is 
the most explicit obstruction to the equality  of $NP$ with $P$ that one
can imagine. Similarly, an $O(1)$-size  program  for a one 
way function is an explicit obstruction for every $n$ and $m=\poly(n)$ 
after there is a final proof 
that it is indeed a one way function, i.e., when it is accompanied by
such a final proof  (without  it it is not even an 
obstruction, as already pointed out). 
But we are assuming here that the approach has not been
so lucky as to come with a final proof right in the first step. That is, 
the new obstruction in this first step is not a 
final proof or  accompanied by a final proof
but only a vehicle towards the final proof
as in the proofs of the earlier nonuniform lower bounds.

\noindent {\bf II. (The strong flip: theoretical feasibility of discovery) 
[optional]:} 

\noindent (a) Give a $\poly(n)$  time algorithm for deciding,
for any given $n$ and $m< 2^n$,
if there exists such a new  obstruction, and constructing the specification of
one if one exists. This means the problem of discovering an obstruction
belongs to $P$, and hence, is theoretically feasible (``easy'').
Formally, 

\begin{defn} \label{defnobsstrongexp}
We say that an explicit family of obstructions
(Definition~\ref{defnobstexplicit})
is {\em strongly explicit} if 
there is a discovery algorithm that given any $n$ and $m < 2^n$ 
decides in $\poly(n)$ time if ${\cal O}_{n,m}$ is nonempty
and, if so, produces in this much time specification $[O_{n,m}]$ of
an obstruction $O_{n,m} \in {\cal O}_{n,m}$. (Note independence of the
running time from $m$.)

We say that a strongly explicit family is {\em extremely explicit}
if, in addition,  the obstructions are easy to decode; i.e., 
given the specification of a new obstruction $O_{n,m} \in {\cal O}_{n,m}$
and a small $C$ of size $m$ the associated decoding algorithm 
produces in $\poly(n,m, \bitlength{O_{n,m}})$ time a counterexample $X$ for 
$C$. 
\end{defn}

Strong explicitness  means that the problem of discovering 
an obstruction (or rather its specification) belongs to $P$, and hence,
is theoretically feasible (``easy''). 

\noindent (b) Also state precise formal 
mathematical hypotheses $MH_{s+1}$, ..., MHr (in addition to the
hypotheses MH1,... , MHs in the first step) assuming which 
the problem for discovering  obstructions belongs to $P$ and 
justify why the $P\not = NP$ conjecture should not stand in the way of 
proving these mathematical hypotheses.

By a reasonable formalization of this step II of the flip,
we mean precise mathematical formulation of $MH_{s+1}$, ..., MHr,
with a reasonable justification for  why the $P\not = NP$ conjecture 
should not stand in the way of their proofs.
By  implementation of this step II, we mean 
formally proving $MH_{s+1}$, ..., MHr (in addition to MH1, ..., MHs in
the  step I)  and thereby proving that 
the problem of discovering  obstructions belongs to $P$.

Again the $P\not = NP$ conjecture seems to stand in the 
way of this step at the surface,
because the conjecture  says discovery is hard, whereas 
this step says  discovery (of its own proof) is easy.

\noindent {\bf III. (Practical feasibility):} 
Using the ``easy'' criterion for recognizing an obstruction 
in step I, prove:

\noindent {\bf OH (Obstruction Hypothesis):} If 
$m=\poly(n)$, then for every sufficiently large $n$, there exists 
such a short obstruction $O_{n,m} \in {\cal O}_{n,m}$. 


If we also have an ``easy'' procedure for discovering an obstruction 
in step II, then this amounts to  proving that this 
procedure always discovers and constructs the specification of 
an obstruction $O_{n,m}$  successfully when $m=\poly(n)$--i.e., the discovery
procedure works correctly (successfully). So this step amount to 
proving correctness of the discovery program.

\begin{defn} \label{dexpproof}
We  call a proof of the $P\not = NP$ conjecture an {\em explicit proof}
based on the given notion of obstructions if 
it is based on the flip and carries out (i.e. implements) 
step I, a {\em strongly explicit
proof} if it carries out both steps I and II, and an 
{\em extremely explicit proof} if
in addition, the obstructions are easy to decode.
\end{defn}

Explicit and strongly explicit proofs 
based on proof-certificates that are easy to verify and discover make sense
in  a large class of problems in mathematics, 
not  just in complexity theory as here. See [GCTflip] for a 
detailed discussion of explicit proofs in mathematics (they were
referred to as $P$-verifiable and $P$-constructible proofs in the earlier 
version of GCTflip). 
We shall not worry about this wider issue  here.

We say that the flip, and hence, {\em the complexity barrier is (reasonably)
formalized}
once OH and MH1,..., MHr are formally stated with a reasonable 
justification for why they should hold and why 
the $P\not = NP$ conjecture should not
stand in the way of the proofs of MH1 to MHr. 
The justification for why the $P\not = NP$ conjecture should not stand in
the way of proving OH, once MH1, ..., MHr are proved, is given below.
We say that {\em the complexity barrier for verification is formalized} 
once OH, MH1, ..., MHs are formally stated with reasonable justification,
and {\em the complexity barrier for discovery is formalized}
once, in addition, $MH_{s+1}$, ..., MHr 
are formally stated with reasonable justification.

We  say that OH becomes {\em theoretically feasible}
once the steps I and II have been carried out (implemented)--and thus 
MH1, ..., MHr are proved--since
the associated problems of verifying and constructing an obstruction 
then become theoretically feasible, i.e., belong to $P$--in other words, 
the theoretically infeasible task of proving IOH is transformed 
into a theoretically feasible task of proving OH, and 
however hard the task of proving OH may be mathematically,
the $P\not = NP$ conjecture  can be expected not to stand in the way
of its proof.
Hence, the name of this strategy {\em the flip}:
from the infeasible to the feasible. 
We also say that the {\em complexity barrier for verification} 
is crossed once step I is carried out, and the {\em complexity barrier
for discovery} is crossed once step II is carried out, and the 
{\em complexity barrier is crossed} once both steps I and II are 
carried out. We also say that the approach becomes {\em theoretically
feasible} once the complexity  barrier is crossed.

Of course, theoretical 
feasibility of OH  does not automatically  imply that the task of 
actually proving it would be practically feasible.
For that, it is also necessary
that the algorithms for verifying and discovering an obstruction in
steps I and II have a simple enough structure that can be used for
proving OH in step III; i.e.,  for proving correctness of the discovery 
procedure. If not, they have to made simpler and simpler until
this becomes practically feasible.
Theoretical feasibility of the approach   basically
means that the $P\not = NP$ conjecture should 
no longer stand in the way of the 
approach, and hence, there is no philosophical barrier to 
practical feasibility, though transforming  a theoretically
feasible  approach into a practically feasible approach  can be  a challenging
mathematical problem in itself.

Step II above  has been labelled optional, because
there may exist a  proof of the $P\not = NP$ conjecture
along the  lines of the flip  that  takes substantial
short cuts in this step.
For example,  a proof technique may simply be lucky and come 
up with a right guess for the obstruction family (since it is meant to be
carried out by humans and not machines). 
But substantial short cuts in step I seem difficult. After all, if the proof
technique cannot even recognize easily what it is searching for, then
there seems to be   a serious problem.

\begin{sidewaysfigure} 
\begin{center}
\psfragscanon
\psfrag{pi3}{{\small $\Pi_3$}}
\psfrag{pi2}{{\small $\Pi_2$}}
\psfrag{P}{{\small $P$}}
\psfrag{NP}{{\small $NP$}}
\psfrag{T}{{\small $T$}}
\psfrag{inpi3}{{\small $DP(IOH) \in \Pi_3$}}
\psfrag{inNP}{{\small $DP(OH) \in NP$}}
\psfrag{inP}{{\small $DP(OH) \in P$}}
\epsfig{file=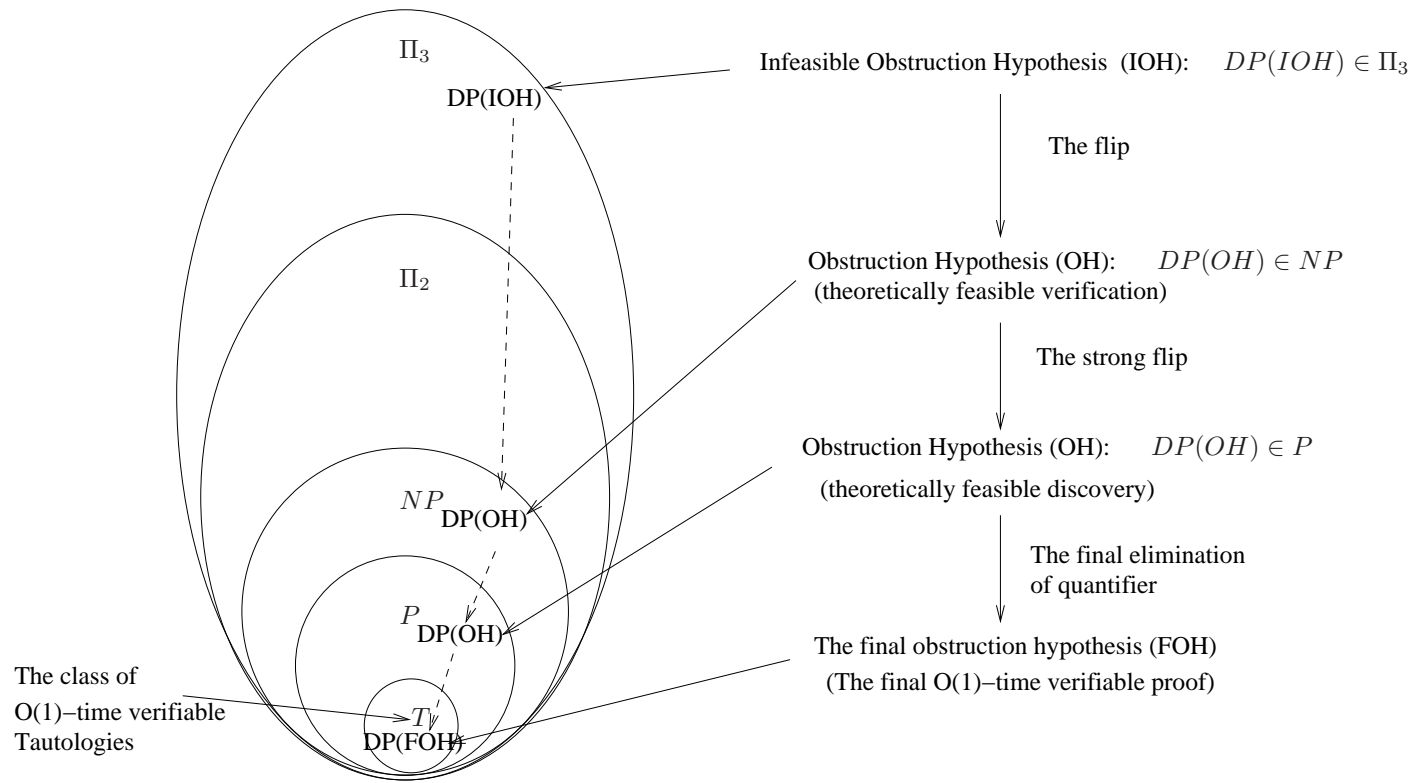, scale=.8}
\end{center}
      \caption{The flip and the polynomial time hierarchy}
      \label{fig:IOHcomplexityintro}
\end{sidewaysfigure}

Here is  another way to look at the flip 
(cf. Figure~\ref{fig:IOHcomplexityintro}). 
For that let us reexamine IOH. 
The problem of deciding,  given $X$ and $C$, if  $f(X) \not = f_C(X)$
belongs to $P^{NP}$.
Since there are two alternating layers of quantifiers in (\ref{eqiohp}),
the problem of verifying it--i.e. the problem of deciding 
if  a trivial obstruction exists for given $n$ and $m$ specified
in unary--belongs to $\Pi_2^{NP} \subseteq \Pi_3$.
We refer to this as the decision
problem associated with IOH and denote it by $DP(IOH)$.
When $m=\poly(n)$,  IOH  is expected to be a tautology,  though
we do not know that yet, and hence, the answer to this decision
problem  would then be always yes.
But for general $m$, not just $m=\poly(n)$, it is a hard decision
problem that  lies  high in the complexity hierarchy
(cf. Figure~\ref{fig:IOHcomplexityintro}) and this is another reason for
referring to IOH as theoretically infeasible.
Let $DP(OH)$ be  the decision problem 
associated with OH, by which we mean the problem of deciding existence of 
a (new) obstruction for given $n$ in unary and $m$  in binary 
(instead of in unary as above--which makes the problem harder). 
It belongs  to $NP$
after execution of the step I and to $P$ after the step II.
Once this  $P$-verifiable OH is proved, it is reduced to a tautology
(FOH: Final Obstruction Hypothesis), which can be verified in 
$O(1)$ time. This gives us the final $O(1)$-size proof.
Thus the basic idea of the flip is to systematically 
eliminate the quantifiers in IOH  and reduce the complexity of the
decision problem associated with 
the  obstruction hypothesis until the obstruction hypothesis  is finally
reduced to  an $O(1)$-time verifiable  tautology;
cf. Figure~\ref{fig:IOHcomplexityintro}. 

But this extremely simple strategy is also deeply paradoxical.
Because it may be conjectured that  $DP(IOH) \in \Pi_3$ 
above is hard, i.e., does not belong to $\Pi_2$ or $\Sigma_2$.
Though this problem may not be $\Pi_3$-complete, we can still call
it a universal problem, because the $P$ vs. $NP$ problem and 
hence IOH is a universal statement about mathematics. 
So to make the flip work, the approach has to come with a new 
notion of an obstruction so that for $m=\poly(n)$ the new OH is equivalent 
to IOH logically (both becoming tautologies),
and yet the decision problem for the 
new OH for general $m$ 
should belong to $NP$ for the flip and $P$ for the strong flip. 
Furthermore, the new OH is also a universal statement
since it is meant to imply $P \not = NP$.
Assumming that the polynomial time hierarchy does not collapse,
or more specifically, that $\Pi_3 \not = NP \not = P$, such reduction
of a universal statement in $\Pi_3$ to another in $NP$ and then
to $P$ may seem impossible.
That is, as the flip starts moving from IOH towards its final 
destination FOH through the polynomial time hierarchy as shown
in Figure~\ref{fig:IOHcomplexityintro},
it may seem as if  it may bring about a collapse of this 
hierarchy in its wake  thereby disproving the very conjectures 
it is meant to prove.   So once again we see the self-referential
nature of the  complexity  barrier which is saying here 
that the main  barrier towards the $P$ vs. $NP$ problem is the problem itself.

For mathematical problems which are not universal, theoretical feasibility
can be meaningless. For example, consider 
a trivial  assertion $\forall A, B: A + B = B + A$ where $A$ and $B$ 
are $n\times n$ matrices.
The problem of brute force verification of this assertion 
belongs to $\Pi_1$, but that is no indication of its difficulty,
and we cannot say that it is a theoretically infeasible hypothesis.
Indeed, it is well known
in proof theory that the logical complexity of an assertion, i.e., the
number of alternating quantifiers in it, is in general no index of its
mathematical difficulty. In the context of the $P$ vs. $NP$ problem 
theoretical feasibility of elimination
of the quantifying variables in IOH
is  a fundamental  issue only because of its universality and because 
those alternating quantifiers are intimately related to the 
the alternating quantifiers used to define the conjecturally 
noncollapsing  polynomial-time hierarchy.
For mathematical
problems that are not universal, e.g. the Riemann hypothesis,  practical 
feasibility is all that matters. Because however hard they may be, they do not
say that discovery is hard.

It may be noticed  that  proving a superpolynomial 
lower bound for $f(X)$  via the flip 
is not very different from proving a superlinear 
lower bound, because the
heart of the flip, consisting of the steps I and II, is the same whether
$m$ is superpolynomial or superlinear, and only the final step III differs.
In other words, on any approach based on  the flip, 
modest as well as  strong lower bounds
lie on the other side of the complexity barrier, and 
the main difficulty in getting to them--the complexity barrier 
(steps I and II)--is the same. 
One may wonder if there is any other gradual way towards the $P$ vs. $NP$
problem on which
we could meet  modest lower bounds much before strong lower bounds without
having to cross  the formidable complexity barrier first.
But in view of the universality of the complexity barrier, it is 
plausible that this is a universal phenomenon. That is, the same thing
may well happen on any approach towards the $P$ vs. $NP$ problem. 

The  flip can also be formulated
in the theory of pseudorandom generators. For that we  fix 
an explicit $NP$-computable pseudorandom generator 
$g$ that conjecturally fools 
small  circuits (assuming which  $P \not = NP$). We  then express 
this property as an IOH for $g$ and associate a decision problem with it.
But it will be seen that this decision problem lies even higher in
the complexity hierarchy than the IOH for the $P$ vs. $NP$ problem; 
specifically it lies  in $\Pi_3^{BPP}$. Hence
the corresponding flip, which begins at this IOH  and runs towards FOH for $g$ 
through the complexity hierarchy, is harder.
In other words,  this says  that the theory
of pseudorandom generation (derandomization)
is in general  harder than the theory of lower bounds, and hence,
hardness should be proved first  and then transformed into
pseudorandom generation  as per the hardness vs. randomness 
principle \cite{nisan}. This is why we focus on lower bounds in GCT at present.

The flip considered so far  is {\em global} because it is 
based on a global  obstruction of $\poly(n)$ size 
that acts as a proof certificate of hardness for all small
circuits. We may also consider a  {\em local  (strong) flip}
based on a {\em local (strongly) explicit 
obstruction}, by which we mean 
a table specifying a counterexample $X$ for each small circuit $C$ of size 
$m$ whose each row can be verified (constructed) in $\poly(n,m)$ time. 
\ignore{It may be conjectured that the local strong flip is possible in 
the context of the $P$ vs. $NP$ problem: i.e. for
any $NP$-complete $f(X)$, there exists an algorithm that 
produces a counterexample $X$ for 
any circuit  $C$ of $m=\poly(n)$ size in $\poly(n,m)=\poly(n)$ time. 
We shall refer to this as the {\em local strong flip conjecture} for
the $P$ vs. $NP$ problem--it is evidently much stronger than the nonuniform
$P\not = NP$ conjecture.}
The local flip  crosses the  complexity 
barrier {\em locally} in the sense that a 
locally  explicit  obstruction is fundamentally different and better 
than a trivial obstruction. Better locally but not  globally
because  verification 
of the whole locally   explicit obstruction still
takes exponential time. The same  for a local strongly explicit obstruction.
In this sense, the global flip is stronger than the local flip.
The global   flip also implies   the local  flip if there is a polynomial 
time decoding algorithm, since then 
from a  global  obstruction one can get 
a local  strongly  explicit  obstruction easily. 
The $P\not = NP$ conjecture is not expected to 
stand in the way of this decoding since it simply amounts to spelling out 
what the approach has been doing implicitly anyway.
Hence intuitively the global (strong) flip subsumes the local (strong) flip.
A local strong flip is used in the proof of the $P\not = NC$ result 
without bit operations described below (Section~\ref{swithoutbit}).
But it does not seem 
effective for the harder separation problems in view of the algebraic 
degree barrier discussed there.

The complexity barrier and the flip for other problems such as 
the $P$ vs. $NC$ and $\#P$ vs. $NC$ problems are analogous.
Though these problems  are not  universal, 
we  assume that the complexity barrier is 
meaningful for them  because we look at them as  variants
of the $P$ vs. $NP$ problem.

So far we have been assumming that the underlying field of 
computation is the boolean field. But the $P$ vs. $NP$ and
$\#P$ vs. $NC$ problems can also be defined over an arbitrary ring or
field, and we are  specifically  interested in 
the rings and fields of characteristic zero such as $Z$, $\Q$ and $\C$. 
Let us now explain what we mean by the (nonuniform) $\#P$ vs. $NC$ and 
$P$ vs. $NP$ problems in characteristic zero. 
By the nonuniform (characteristic zero) version of the $\#P$ vs. $NC$ problem
we mean  the permanent vs. determinant problem \cite{valiant}. 
It  is to show that $\perm(X)$, the permanent of an $n\times n$ variable 
matrix $X$, cannot be represented 
linearly  as $\det(Y)$, the determinant of 
an $m\times m$ matrix $Y$,
if $m=\poly(n)$, or more generally, $m=2^{\log^a n}$, 
for a fixed constant $a>0$, 
and $n\rightarrow \infty$. 
By linear representation, we mean
the  entries   of $Y$ are  (possibly nonhomogeneous) 
linear functions (over $\C$, $\Q$,  or $Z$)
of the entries of $X$. 
There is an analogous characteristic zero version of
the $P$ vs. $NP$ problem defined in GCT1, where the role of
the permanent is played by an appropriate (co)-NP complete function and
the role of the determinant is played by an appropriate $P$-complete function.
But we shall not worry about it here because all the basic ideas of GCT
are  illustrated in the $\#P$ vs. $NC$ problem in characteristic zero.
In this article we  mostly focus  on characteristic zero     because
the nonuniform lower bounds over $Z$ are weaker implications of the
corresponding ones over a finite field (the usual case), and hence
easier. But since the actual $P$ vs. $NP$ problem
is over a finite field (boolean), the flip over a finite field 
is briefly discussed in Section~\ref{sexpli}, with a  detailed discussion 
of the problems that arise in this context  postponed to   GCT6 and 11. 

The flip can also be defined for the $P$ vs. $NP$ and $\#P$ vs. $NC$ 
problems over fields of characteristic zero such as $Q$ or $\C$ and for
algebraically closed fields of positive characteristic such as $\bar F_p$, 
the algebraic closure of the finite field $F_p$, $p$ prime.
All  definitions in the flip 
can be lifted to this setting in the obvious way. We  only give
one critical definition as an illustration:

\begin{defn} \label{ddecodeQ}
Assume that the underlying 
field is $Q$. In this case we say that an
obstruction family is {\em easy to decode} 
if there is a decoding algorithm which given the specification of 
an obstruction $O=O_{n,m}$ and a small circuit $C$ over $\Q$ of size $m$ 
(i.e. whose whole description is of bitlength $m$)
produces in $\poly(n,m, \bitlength{O})$ time a counterexample (local
obstruction) $X$ for 
$C$ whose description has  $\poly(n,\bitlength{O})$ bitlength, 
independent of $m$ as long as it is $\le n^{\log n}$ (say). If 
$\bitlength{O}$ is $\poly(n)$ the bitlength of $X$ is then $\poly(n)$
independent of $m$. In other words, smallness
of  obstructions independent of $m$ is to be preserved
in the process of decoding.

Definition over $\bar F_p$ is analogous.
\end{defn}

The flip is the defining strategy of GCT. Indeed, GCT may be abstractly
defined as any (geometric) approach to the $P$ vs. $NP$ and related 
problems based on the flip, i.e., explicit construction of obstructions
(for a precise meaning of the phrase geometric see Section~\ref{scharsym}).
By based on the flip, we mean it uses the flip as a basic guiding principle
and goes for ``more or less'' explicit construction of obstructions, 
though it does not have to be completely explicit, i.e., 
some deviations and short cuts are fine and also inevitable as long as 
the spirit  is maintained.
The present (concrete) GCT is one such  approach.
It proposes  a scheme for implementing the flip for the $P$ 
vs. $NP$ and related problems via algebraic geometry, representation
theory and the theory of quantum groups.
This turns out to be a nontrivial affair, as to be expected, because of the
counterintuitive and paradoxical  nature of the flip as explained above.

But before we enter into it, let us 
ask ourselves if the flip, i.e.,  explicit construction, the root 
cause of all difficulties in GCT, is really necessary.
Because, to solve  the  lower bound problems under consideration
we only need to prove existence of obstructions, we do not have to
construct them,  and moreover,
the flip is only the most obvious and natural way to cross
the complexity barrier. 
There may exist  nonobvious and 
unnatural (e.g. nonconstructive) ways to cross this barrier.
At present,  we only know 
that one class of nonconstructive techniques 
will not work, namely the ones with probabilistic flavour, which
are ruled out by the natural-(probabilistic)-proof-barrier \cite{rudich}. 
But the probabilistic  proofs constitute  only a 
small subarea of nonconstructive proofs, namely the ones 
which are of a  {\em nonspecific} kind, by which we mean nonconstructive
proofs which apply to most functions or structures under consideration.
Vast areas of
nonconstructive proofs in  mathematics, such as the ones  in algebraic 
geometry and representation theory, are of fundamentally different nature
than probabilistic proofs: they are nonconstructive yet 
{\em very specific};
i.e., they  apply to only very specific functions or structures
under consideration. The natural-(probabilistic)-proof-barrier 
does not apply to
such  nonconstructive but specific proof techniques for the reasons 
explained in Section~\ref{scharsym}. 
Indeed, one of the concrete lower bounds of GCT to be discussed below
(Theorem~\ref{tmathform}) has a nonconstructive   proof
which applies to only very specific  functions (generalized permanents) 
and which 
bypasses  the natural-proof, relativization, and 
algebraic-degree (algebrization) 
barriers simultaneously  in this   setting. 
This gives  concrete evidence that
explicit constructions are  not needed to bypass these local barriers, that
they may be bypassed by nonconstructive proof techniques as well, and that 
there is no reason why the universal complexity barrier too could not be
crossed by such nonconstructive techniques.

But the mathematical evidence and arguments 
in [GCTflip] suggest that such unnatural ways to cross the complexity barrier
may  be  even harder than  the natural ways as one may intuitively expect. 
For example, the nonconstructive proof of the lower bound mentioned above
is not much different or easier than the explicit proof, and in fact
a bit harder.
We do not go any further into such  arguments  here.
At a concrete level, 
all we can say is that we are going the natural  way because there 
is no  concrete  alternative at present.
After all, as pointed out in Section~\ref{scomplexity}, 
any approach to the $P$ vs. $NP$ problem has 
to meet at least the criteria A, B, C, and  D stated there. 
We also  stated there why GCT meets these criteria.

\ignore{The $P \not = NC$ result without bit operations [GCTlocal] in GCT
(cf. Section~\ref{swithoutbit}) addresses criterion A.
The mathematical form of the $\#P\not = NC$ conjecture [GCT1,2,6] 
(cf. Section~\ref{smathform})   addresses criterion B.
The main result [GCT6,7,8,11] of GCT  (cf. Sections~\ref{sgct6} and 
\ref{sexpli})  gives a conjectural
scheme supported by mathematical evidence to cross the complexity barrier
via the flip for  the nonuniform
$P$ vs. $NP$ and $\#P$ vs.
$NC$ problems over fields of characteristic zero and also 
over the finite field (the usual case). This
addresses criterion C. }

But, at present,  we do not have any  concrete   alternative to GCT
which addresses A, B, C  and D, or for that matter even  one of them,
and hence  there is not much that we  can
do but follow  the most natural direction.

We now briefly  describe  the   concrete results of GCT 
which address A, B, C, and D.

\section{A: The $P\not = NC$ result without bit operations} \label{swithoutbit}
We  begin with 
a  special case of the $P \not = NC$ conjecture [GCTlocal],
which we are calling the $P\not = NC$ result without bit operations,
and which can be considered to be the
first concrete lower bound result of GCT. It was  proved using the weaker
local form of the flip much before the (global) flip was formalized
in [GCT6,GCTflip].
It  says that:

\begin{theorem} \label{tmax}
The $P$-complete max-flow problem
cannot be solved in $\polylog(N)$ parallel time using $\poly(N)$
processors in the PRAM model without bit operations,
where $N$ denotes
the total bit length of the input, not just the number of input parameters.
\end{theorem}

The model here is the usual PRAM model with arithmetic $+,-,*$, comparison
and branching operations, but no bit operations.
It  is quite realistic and natural
since it includes virtually all known parallel algorithms for algebraic 
and weighted optimization problems. Hence, this lower bound meets 
the criterion A.

Its proof is based on classical algebraic geometry, though 
the result itself can be stated in purely 
elementary and combinatorial terms; no elementary or combinatorial 
proof  is known so far.
This proof technique is  locally strongly
explicit in the sense that  for each branching program $C$  (i.e. 
a circuit with arithmetic, branching and comparison operations) 
of depth $\log^a N$ and size $\poly(N)$, or even 
$2^{\log^a N}$, for some positive constant $a$,
it produces in $\log ^b N$ time using $2^{\log ^b N}$ 
processors, $b=O(a)$, 
an explicit   counter example  $X$ such that the output of $C$
on $X$ is different from the result of  the maxflow (decision) problem.
This counter example $X$ is then a local 
proof-certificate of hardness (obstruction)
against $C$. For this reason we shall refer to this proof technique 
as local GCT.
(The definition of local explicitness here is a bit relaxed 
in the sense that we are considering circuits of size $2^{\polylog(N)}$ 
instead of $\poly(N)$.)  

Local explicitness was not  specifically mentioned 
in GCTlocal. But it is easy to see \cite{GCTpramappend}  that 
the proof technique there actually
produces a locally 
explicit proof certificate of hardness, after a little postprocessing,
though the goal was only
to prove its existence (which is why explicitness was not mentioned).
In other words, the proof technique essentially produces such a 
proof-certificate explicitly whether one cares for it or not; cf. 
\cite{GCTpramappend} for details.

Criterion A and local explicitness of the proof technique is 
the fundamental difference between
this lower bound result and the earlier ones, such as 
the lower bounds for constant depth circuits \cite{sipser}, monotone circuits 
\cite{monotone}, or 
algebraic decision trees \cite{benor}, which do not meet criteria A,
as far as we can see, and whose proof techniques are inherently 
nonconstructive. For example, the random restriction method \cite{sipser}
for proving lower bounds for  constant depth circuits gives a randomized 
fast parallel algorithm to produce  for each such
circuit $C$ a local counter example $X$ on which it differs from
the parity function. But the best deterministic algorithm to produce 
such a counter example, based on  derandomization of the switching
lemma \cite{agarwal},
takes sequential polynomial time. Locally strongly explicit in this
setting would mean computable by a constant depth circuit
(or in a relaxed sense, 
say $\log \log n$ depth circuit of $2^{\polylog(n)}$ size). 
Similarly, the proof of the lower bound for sorting in
the  algebraic-decision-tree 
model \cite{benor}  inherently requires
exponential time  to produce a counter example, 
so also the proof of the lower bound for monotone circuits \cite{monotone}.
It may be conjectured that there are
(locally strongly) explicit proofs 
for the constant-depth or  monotone circuit lower bounds. 
At present this is  open. But the main point is that such  explicit 
proofs are not  required to prove these lower bounds,
whereas, 
the only way we can prove the  the $P \not = NC$ result without bit operations
at present is essentially via locally explicit construction.

\noindent {\em Remark:} Assuming  a stronger lower bound 
\cite{hastad}  for 
constant depth circuits,  a counter example 
against a constant depth circuit can be easily constructed  in 
$\polylog(N)$ time using
$2^{\polylog(N)}$ processors. And similarly assumming a stronger form
of the $P\not = NC$ result without bit operations, a counter example 
against a branching  circuit of $\polylog(N)$ depth
and $2^{\polylog(N)}$ size can be constructed  by a branching circuit of
$\polylog(N)$ depth and $2^{\polylog(N)}$ size.
But what matters here is not 
how quickly obstructions can be constructed 
once  a lower bound is proved, but rather how quickly the obstructions
that are  used to prove the lower bound  can be constructed.

This discussion suggests that local GCT 
is fundamentally different from the probabilistic techniques to which
the natural proof barrier applies and also the uniform techniques to which
the relativization barrier applies.

But that is not enough for  the harder
lower bound problems in Figure~\ref{fig:complexityclasses}
because  of the third local barrier in complexity theory
that was pointed out in GCTlocal towards the end of its Chapter 7. We shall 
call it  the {\em algebraic degree barrier}, though it was not
given any name there. This  says that 
any low degree or degree based proof technique, such as the one there,
which associates with a computation algebraic 
objects (polynomials, varieties, etc.) 
and then reasons solely on the basis of the degrees of those 
objects will not work for unrestricted 
fundamental separation problems. 

This suggests that a local flip may not be good 
enough to prove lower bounds for unrestricted circuits
that are  allowed to carry out all field operations
unlike the restricted circuits for which lower bounds can be proved by 
exploiting absence of fundamental field 
operations. Because an unrestricted circuit is such a complex and 
unstructured mathematical object that 
its  only algebraic invariant  that a local flip that 
works on each circuit separately 
can use effectively is the  degree, and the algebraic degree
barrier says that such degree based reasoning would not work.

GCTlocal 
also suggested in its chapter 7 
an idea for crossing the algebraic degree barrier
and this limitation of the local flip: namely, associate with complexity 
classes algebraic varieties with group actions that capture the symmetries
of computation and then reason globally 
on the basis of the deeper representation theoretic 
structure of these varieties rather than just their degrees.
Algebraic varieties with group action have been  studied intensively
in geometric invariant theory \cite{mumford} for over a century.

\section{Characterization by symmetries} \label{scharsym}
This   motivated the approach to  fundamental lower bound problems 
via geometric invariant theory that was initiated   in GCT1 and 2. 
We  now describe the basic ideas of this approach  focussing on the
permanent vs. determinant problem. 
It   begins with an 
observation [GCT1] that the permanent and determinant are 
{\em exceptional} polynomial functions, where by exceptional we mean
they are  completely characterized by  symmetries in the 
following sense.

Let  $Y$ be a variable $m\times m$ matrix.
Let $\sym^m(Y)$ be the space of homogeneous forms of degree $m$ 
in the $m^2$ variable entries of $Y$. Then by classical 
representation theory  $\det(Y)$ is the only form in $\sym^m(Y)$ 
such that, for any $n \times n$ invertible 
matrices $A$ and $B$ with $\det(A)\det(B)=1$, 

\noindent  {\bf (D):} $\det(Y)=\det(A Y^* B)$,

where $Y^*=Y$ or $Y^t$.  This is so for any underlying field 
of computation. 
Thus $\det(Y)$ is completely characterized 
by its symmetries and hence is  exceptional. We shall refer 
to this characteristic property of the determinant as property (D)
henceforth. 

Similarly by classical representation theory  $\perm(X)$ is the only form 
in the space $\sym^n(X)$ of homogeneous 
forms of degree $n$ in the entries of $X$ such that, 
for any diagonal or permutation matrices $A,B$,

\noindent  {\bf (P):} $\perm(X)=\perm(AX^*B)$, 

where $X^*=X$ or $X^t$  with obvious constraints 
on the product of the diagonal entries of $A$ and $B$ when they are diagonal.
This is so for any underlying field of computation of characteristic 
different than two.
Thus $\perm(X)$ is also
completely characterized   by its symmetries and hence is
exceptional.
We shall refer 
to this characteristic property of the permanent as property (P)
henceforth. 

The special functions that play the role of the determinant and
the  permanent in the $P$ vs. $NP$ problem in GCT1 are similarly
characterized by their symmetries in a slightly weaker sense but which
is good enough for our purposes. For the sake of completeness, 
let us just focus on the permanent vs. determinant problem here.

The Useful Property (UP), using the  terminology of \cite{rudich}, 
of the permanent  that 
GCT plans to use to prove a superpolynomial lower bound is:

\noindent {\bf UP:} (1) It is $\#P$-complete, and (2) It is characterized 
by its symmetries.

Here (1) is just meant to ensure that the permanent is hard--anything 
that ensures hardness will do here. The
property that really drives GCT is (2)--characterization by symmetries.

Let us now see how symmetries can help us tackle the
local barriers; we will turn  to the universal complexity barrier later.

The relativization 
barrier \cite{rudich} is not really relevant in the permanent vs. determinant  problem
since in its statement 
$NC$-computation has already been algebrized in the form of the determinant
and by now it has been  accepted in complexity theory
that  this barrier does not apply to proofs 
that work over algebrized forms of computation, e.g., the proof of
the $IP=PSPACE$ result. 
Similarly the algebraic 
degree [GCTlocal] (and also algebrization \cite{aaronson}) barrier is also not 
relevant since the permanent and   determinant 
have the same degree and hence a proof technique that 
relies just on degrees would not be able to distinguish between them.
In other words, the relativization, algebraic degree and algebrization
barriers are automatically bypassed in this problem 
by the very nature of its statement. So 
the only local barrier that really matters here 
is the natural proof barrier. Let us now see 
how symmetries can help in bypassing  this barrier.
Towards that end, we need a few definitions.

Fix any  polynomial $p(X)$ in $\sym^n(X)$. It 
can be described by giving all its  coefficients.
If the base field of computation is $F_p$, the finite field of $p$ elements,
the total bit length of this description is 
bounded by $\bitlength{p} D$, where 
$\bitlength{p}$ denote the bitlength of $p$ and
$D={n+n^2-1 \choose n-1}=2^{O(n^2)}$ 
is the total number of  monomials in the entries of $X$ with degree $n$.
We let $N=\bitlength{p} D$ be this bound on the bitlength.
If the base field is $\Q$, we stipulate that each coefficient be of atmost 
$b=\poly(n)$ bitlength (we can also let $b=\poly(D)$). In that case 
the total  bitlength of this description is bounded by $b D$, and we let
$N=b D$ in this case. 
We say that a nonsingular $n^2 \times n^2$ matrix $\sigma$ over the 
base field is a symmetry of $p(X)$ if $p(\sigma X)=p(X)$, where 
we think of $X$ as an $n^2$ vector after straightening it, say, rowwise.
We say that $p(X)$  is completely characterized by symmetries if the only 
function in $\sym^n(X)$ whose symmetries contain    the symmetries of 
$p(X)$ is a constant multiple of $p(X)$.

\begin{defn}  \label{tuseofsym}
\noindent (a) We say that a proof technique for the permanent vs. determinant 
problem (or its restricted version)
makes a {\em nontrivial use of symmetries} if it works (i.e., can 
prove a lower bound) only
for  functions $p(X)$ (in place of the permanent) that have nontrivial
symmetries with  explicit description. By explicit description 
we mean  in  terms of $\poly(n)$ 
generators with description of  $\poly(n)$ bitlength. 
By nontrivial we mean the symmetries  do  not consist of just the identity
transformation or its multiples.

\noindent (b) We say that a proof technique  is 
{\em characterized by symmetries} if 
it only works for 
$p(X)$ which are completely characterized by symmetries
with explicit description. 
\end{defn} 

The following is an analogue of Definition~\ref{drigidcomplexity}
in this setting.

\begin{defn} \label{drigid}
We say that a proof technique 
is {\em nonrigid} if the total number of  $p(X)$ in $\sym^n(X)$ 
for which it works is $\ge 2^N/poly(N)$, and 
{\em mildly rigid} if this number 
is   $ \le 2^N /N^c$ for some constant $c>0$, or
more generally, $\le 2^N/2^{\poly(n)}$.
We say  it is {\em rigid} if  the number is $\le 2^{\epsilon N}$,
for some $0 \le \epsilon < 1$, {\rm strongly rigid} if the number  is
$\le 2^{N^\epsilon}$, and {\em extremely rigid} if the number  is
$\le 2^{\poly(n)}$.
\end{defn} 

Just as in  Section~\ref{scomplexity},
probabilistic proof techniques to which the natural proof
barrier \cite{rudich} applies are nonrigid, and 
if a proof technique is mildly rigid it bypasses this
barrier, since it violates the largeness 
criterion in \cite{rudich}  with $N$ playing the role of the truth table size 
there.

\begin{prop}  \label{prigid}
Let the base field  be $\Q$ or a finite field $F_p$,
with $p$ large enough, of $\poly(n)$ bitlength. 
A proof technique over such a base field 
that makes nontrivial use of symmetries is 
rigid. A proof technique that is characterized by symmetries is
extremely rigid.
\end{prop} 
(The proof will appear in the full version of this paper.)

Thus a proof technique that makes nontrivial use 
of symmetries automatically bypasses the natural proof barrier
even if it is nonconstructive. 
Most of the proof techniques in  representation theory and
algebraic geometry of varieties with symmetries, i.e., geometric 
invariant theory \cite{mumford}) are nonconstructive. But they
are rigid  because
they make nontrivial use of symmetries, 
in contrast to the nonrigid proababilistic proof techniques to which
the natural proof barrier applies.

As we mentioned above, the useful property in GCT is  UP. Hence, 
Proposition~\ref{prigid} implies

\begin{cor} 
The proof technique of GCT for the permanent vs. determinant problem
is extremely rigid.
\end{cor}
It can be shown that 
this is also the case for the $P$ vs. $NP$ problem. This meets 
criterion D in Section~\ref{scomplexity}.

In Section~\ref{sflip} we defined abstract 
GCT as any geometric technique based on the flip. 
Now we can explain the meaning of the phrase geometric.
Since symmetry is a fundamental geometric concept, 
we say that a proof technique is {\em geometric} 
if it makes a nontrivial use of symmetries and 
is ``more or less'' characterized by symmetries. 
By more or less,  we mean characterization by symmetries
is used as a guiding principle (like the flip)
and some deviations  are fine as long as the spirit is maintained. 
For example, the functions that play the role of the permanent and
the determinant in the $P$ vs. $NP$ problem in GCT1 
are not completely characterized by symmetries. But they are
more or less, enough for our purposes, and also enough to make the
technique extremely rigid. 

The  geometric component of abstract GCT lets it
tackle  the natural proof barrier by Proposition~\ref{prigid}, and  the flip 
component lets it  tackle  the complexity barrier.
But, as we already remarked in Section~\ref{scomplexity},
extreme rigidity implied by  characterization by symmetries 
(Proposition~\ref{prigid}) is 
far more than   mild rigidity that 
suffices to cross the natural proof barrier.
Rather, characterization by symmetries is meant to be used in abstract GCT
for   explicit constructions 
which can then  be used to  cross the universal
complexity barrier via the flip, and these explicit constructions,
as per the Rigidity Hypothesis,
have to be extremely rigid.
In other  words,  
though mildly rigid nonconstructive techniques would 
suffice to cross the local natural proof barrier,
GCT goes for extremely rigid explicit constructions based 
on the characteristic symmetry properties (P) and (D) 
with an eye on the universal complexity barrier and because of
the Rigidity Hypothesis.

Representation theory and 
algebraic geometry enter inevitably into  the study of these properties,
because
to understand symmetries representation theory (of groups of symmetries)
becomes indispensible, and to understand  deeper properties 
of representations  algebraic geometry becomes indispensible. 
Thus geometry in the form of  symmetries naturally  leads to algebraic 
geometry, specifically, algebraic geometry of spaces  with group action,
i.e. geometric invariant theory \cite{mumford}. 

\section{A mathematical form of the  $\#P \not = NC$ conjecture} 
\label{smathform}
To illustrate the power of symmetries, we now give an application of GCT in the
form of a second concrete lower bound--namely a 
mathematical form of the 
$\#P \not = NC$ conjecture in characteristic zero 
(Theorem~\ref{tmathform} below).

For that we need a definition.

\begin{defn} \label{dgenperm}
A polynomial function $p(X_1,\ldots,X_k)$ of any degree
in the entries of $k$ $n\times n$ variable matrices $X_1,\ldots,X_k$  
is called a  {\em generalized permanent} if it has exactly 
the same symmetries as that of the permanent. This means
for all nonsingular $n\times n$ matrices $U_i$ and $V_i$, $i \le k$, 
\[ p(U_1 X_1 V_1,\ldots, U_k X_k V_k)=p(X_1,\ldots,X_k) 
\mbox{\ iff\ } \perm(U_i X V_i)=\perm (X) \quad \forall i.\] 
\end{defn}

Here $U_i$ and $V_i$ 
are as in  the property (P) and the underlying field has characteristic
different than two. When $n=1$, any function of $k$ variables $x_1,\ldots,
x_k$ is a generalized permanent. But for general $n$ there are far fewer 
generalized permanents and their  space  has a highly nontrivial structure.

\begin{obs} \label{oimpli}
Assuming the nonuniform $\#P \not = NC$ conjecture in characteristic zero,
no $\#P$-complete generalized permanent  $p(X_1,\ldots,X_k)$ of $\poly(n,k)$ 
degree can be expressed as an $NC$-computable polynomial function of the 
traces of  $\bar X_i^j$, $1 \le i \le k$, $j=\poly(n,k)$, where 
$\bar X_i=B_i X_i C_i$,  $i \le k$,  for any $n\times n$ complex (possibly
singular) matrices $B_i$ and $C_i$.
\end{obs}
This follows because   $\bar X_i^j$ are  clearly  $NC$-computable.

When $n=1$ and $k$ is arbitrary, this implication is  equivalent 
to the original nonuniform $\#P \not = NC$ conjecture in characteristic 
zero, since then all matrices becomes ordinary variables,
any polynomial in $x_1,\ldots, x_k$ is a generalized permanent,
and a polynomial function of  the traces of  $x_i$'s means any
polynomial in $x_1,\ldots,x_k$. This
cannot be proved unconditionally at present.
But the next case of this implication,  $n>1$ and $k$ arbitrary, can be.
That is, the implication in Observation~\ref{oimpli} holds unconditionally 
for $n>1$ and arbitrary $k$. In fact, 
in this case a stronger result holds. Namely:

\begin{theorem} \label{tmathform}
\noindent {\bf (A mathematical form of the  $\#P \not = NC$ conjecture
in characteristic zero)}
No generalized permanent $p(X_1,\ldots,X_k)$
can be expressed as a polynomial function of the traces of 
$\bar X_i^j$, $1 \le i \le k$, $j \ge 0$,  where 
$\bar X_i=B_i X_i C_i$,  $i \le k$,  for any $n\times n$ complex (possibly
singular)  matrices $B_i$ and $C_i$.
\end{theorem}

The result also holds for any algebraically closed field of 
characteristic different than two.

This result is called a mathematical form because 
first there is no restriction here on the 
computational complexity of
$p(X_1,\ldots,X_k)$ or the polynomial in the traces. 
Thus it is rather in the spirit of the classical result of Galois theory 
which says that a  polynomial  whose Galois group is not
solvable cannot be solved by any number of
radical operations, without any restriction on the number of such
operations. 
Second, the permanent has two characteristic properties: 1) the property 
(P) (mathematical), and 2) $\#P$-completeness (complexity-theoretic). 
The usual complexity theoretic form of the $\#P\not = NC$ conjecture
is a lower bound for  all polynomial functions with 
the  $\#P$-completeness property, whereas the mathematical form 
is  a lower bound for all polynomial functions with the 
property (P).  In other words,  the complexity theoretic form
is associated with the $\#P$-completeness property of the
permanent and the mathematical form with its  property (P).

This result indicates that there is thus a huge chasm
between the two adjacent cases: $n=1$, $k$ arbitrary (the usual nonuniform
complexity theoretic $\#P \not = NC$ conjecture), and 
$n=2$, $k$ arbitrary (its mathematical form above). This chasm 
is nothing but  a reflection of the  complexity barrier.

A {\em weak form} of this result, obtained by letting 
$k=1$ and $p(X_1)$ the usual permanent,
was earlier stated in the IAS lectures in February 2009 \cite{IAS}. 
The article \cite{boris} showed that it is too weak by giving a 
an elementary linear algebraic proof. The mathematical form
does not have an elementary linear algebraic proof.

It has two proofs.
The first  uses only basic  geometric invariant 
theory \cite{mumford}. Basically the proof of the weak form (or rather
its nonhomgeneous version) given 
in \cite{GCT2append} works here as well with obvious modifications;
Bharat Adsul \cite{bharat}
has also independenly found a similar proof. But this  proof is naturalizable;
i.e. it cannot cross the natural proof barrier in \cite{rudich} as 
pointed out in \cite{GCT2append}. 

The second proof, discussed in Section~\ref{sgeomobs} below, 
is not naturalizable, and is based on the global flip  strategy  
for the general 
permanent vs. determinant problem and hence serves as a test case for 
the global flip in a nontrivial special case.
It produces a strongly explicit family $\{O_{n,k}\}$
of  (global) obstructions, 
where each $O_{n,k}$ has a specification 
of  $O(n k)$ bitlength  which   can be verified
and constructed in $O(n k)$ time.
This obstruction is global in the sense
that it is against all $B_i$'s, $C_i$'s and the polynomials in the traces. 

This proof technique makes nontrivial use of symmetries and hence
bypasses the natural proof barrier by Proposition~\ref{prigid}. 
In fact, it can be shown to be
strongly rigid  taking the
underlying field to be $\Q$  (Definition~\ref{drigid}), 
and hence it bypasses the natural proof barrier \cite{rudich} in 
a strong sense.
Since the statement of the mathematical form is algebraic and degree 
independent, i.e., it only relies on the property
(P) and makes no mention  of the degree of $p(X_1,\ldots,X_k)$, 
the   relativization,
algebraic degree and algebrization  barriers are automatically 
bypassed   here in the same sense that they are bypassed in 
the permanent vs. determinant problem as we discussed above.
Hence this proof technique (the global flip)
bypasses the relativization, natural proof and algebraic degree barrier
simultaneously, and thus  addresses criterion B.
In a sense the proof technique is more important here than
the lower bound (Theorem~\ref{tmathform}), which is basically  used as a test
case for the proof technique.

\ignore{
To summarize the discussion so far (Figure~\ref{fig:summary}), the 
$P\not =NC$ result without bit operations in GCT is
a complexity theoretic lower bound proved using  a local strongly  explicit  
technique in contrast to the probabilistic techniques used to prove
constant depth \cite{sipser}  and monotone \cite{monotone} circuit 
lower bounds. The mathematical form of the $\#P\not =NC$ conjecture is a 
mathematical lower bound  proved using a global strongly 
explicit and strongly rigid technique.
The mathematical form  deals satisfactorily with the 
local barriers, but not the main complexity 
barrier, though its proof  provides concrete support for the global 
flip strategy to cross this barrier  in a simpler mathematical setting.

\begin{figure} 
\begin{center} 
\epsfig{file=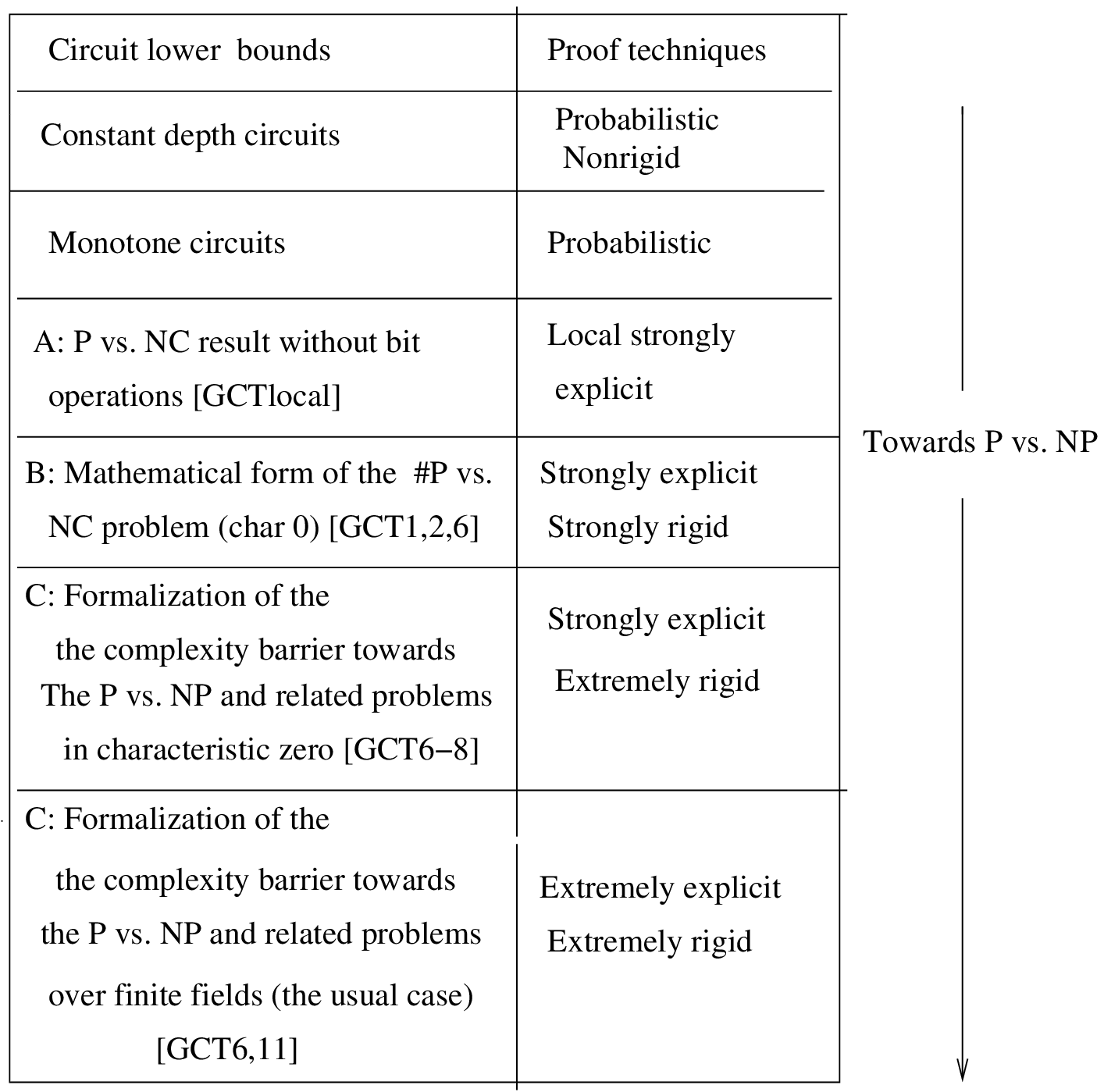, scale=.9}
\end{center}
      \caption{Circuits lower bounds vs. their proof techniques}
      \label{fig:summary}
\end{figure} 
}

\section{Geometric obstructions in characteristic zero} \label{sgeomobs}
Next  we turn to formalization of the complexity barrier via the flip 
and thus address criterion C. 
We describe this 
for the permanent vs. determinant
problem, the story for the $P$ vs. $NP$ problem being  analogous.
We also  point out how a restricted form of the OH that arises 
in  the mathematical form of the $\#P$ vs. $NC$ problem 
(Section~\ref{smathform})
can be unconditionally proved. 
Up to the end of Section~\ref{sgct6},
we assume that the base field
has characteristic zero. We turn to 
the usual finite field case in Section~\ref{sexpli}. 

For  the first step of the flip (Section~\ref{sflip}),
we need  a new notion of obstructions 
that is  fundamentally different than the trivial obstructions.
Such obstructions based on symmetries, which 
we shall call {\em geometric  or representation 
theoretic obstructions}, are  defined in GCT1 and 2.

This is done in two steps.
First, GCT1  associates with 
the complexity class $\#P$ a family $\{X_{\#P}(n,m)\}$ of projective algebraic
varieties, where by a projective 
algebraic variety  we mean  the zero set of a system of 
homogeneous multivariate polynomials with coefficients in $\C$.
We call $X_{\#P}(n,m)$ the class varieties 
associated with complexity class $\#P$.
Each  class variety $X_{\#P}(n,m)$ is   {\em group-theoretic}  in the 
sense that (1) it's  construction is completely characterized (determined)
by the property (P) of the permanent, (2)
the general linear group $G=GL_l(\C)$, $l=m^2$, of $l\times l$ invertible
complex matrices acts on it, and  (3) the group of
the symmetries of the permanent, which we shall
refer to as $G_{perm}$, is embedded in  $G$
as its subgroup in some way. Here action of $G$  means it moves the points 
of the class variety around, just as it moves the points of $\C^l$ around 
by the standard action via invertible linear transformations.

Similarly GCT1  associates 
with the complexity class $NC$ a family of 
$\{X_{NC}(n,m)\}$ of  group theoretic class varieties 
based on the property (D) of the determinant. It also 
has  the action of $G$, and $G_{det}$, 
the group of the symmetries of the determinant, is 
embedded as a subgroup of $G$ in some way.

These class varieties have the property  that
if $\perm(X)$, $\dim(X)=n$, can be represented linearly 
as $\det(Y)$, $\dim(Y)=m>n$, then 

\begin{equation} \label{eqembed} 
X_{\#P}(n,m) \subseteq X_{NC}(n,m).
\end{equation} 

\begin{figure} 
\begin{center} \psfragscanon
\psfrag{xsharp}{$X_{\#P}(n,m)$}
      \psfrag{xnc}{$X_{NC}(n,m)$}
      \psfrag{g}{$G=GL_l(\C)$, $l=m^2$}
      \psfrag{Obstruction}{\mbox{Obstruction}}
\epsfig{file=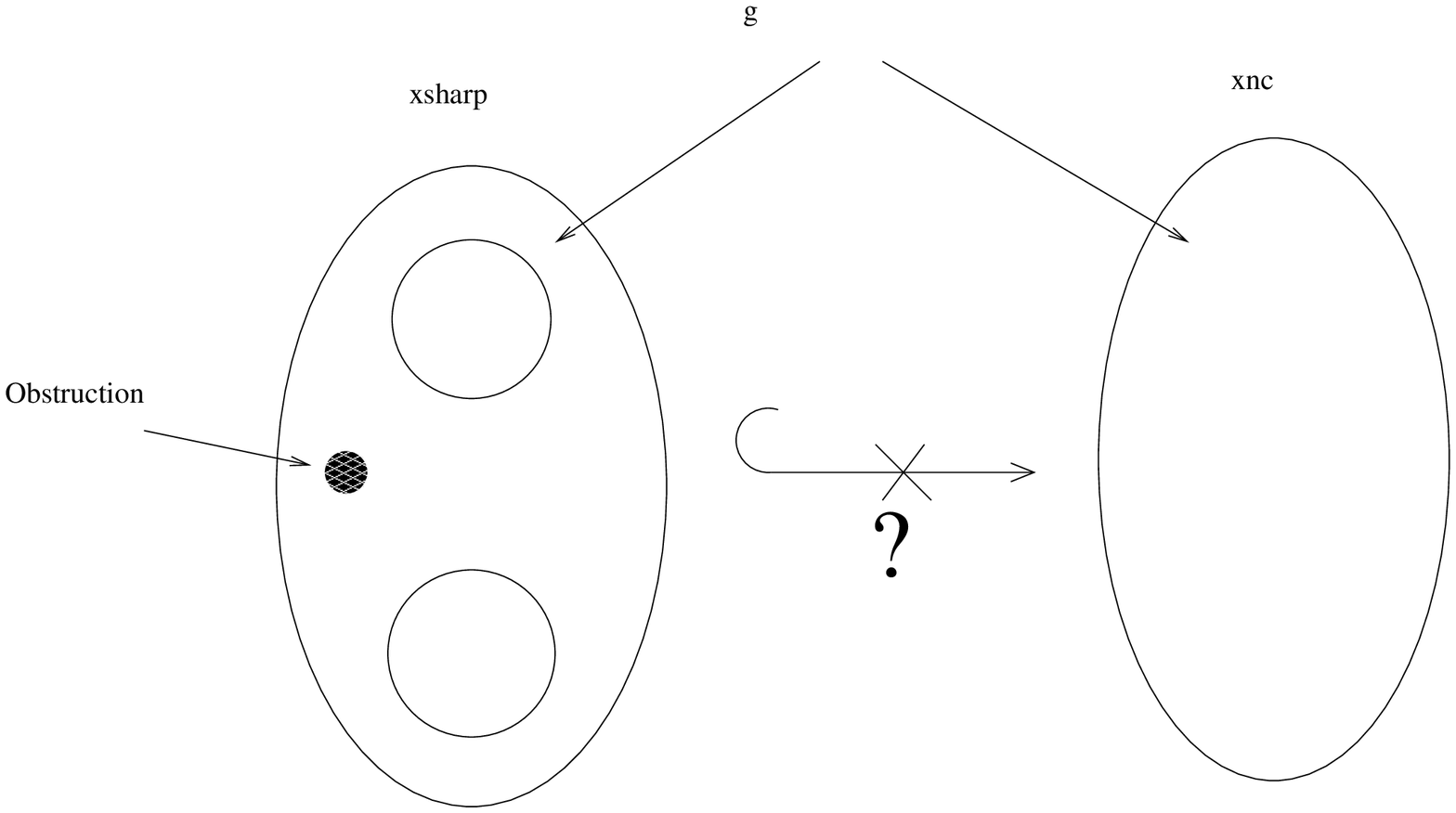, scale=.5}
\end{center}
      \caption{Class varieties}
      \label{fig:class}
\end{figure}

The goal is to show that this inclusion (\ref{eqembed}) is impossible
when $m=poly(n)$ or more generally $m=2^{\log ^ a n}$, $a>0$ a constant.
Geometric obstructions are meant to ensure this;
cf. Figure~\ref{fig:class}. 

To define them  formally, we need to recall some basic representation
theory.  By a classical result of Weyl, the irreducible 
(polynomial) representations of $G=GL_l(\C)$ are in one-to-one correspondence 
with the partitions  $\lambda$ of length  at most $l$, by which we mean 
integral sequences $\lambda_1 \ge 
\lambda_2 \cdots \ge \lambda_k > 0$, $k \le l$,
where  $k$ is called the  length
of $\lambda$. The irreducible 
representation of $G$ in correspondence with $\lambda$ is denoted 
by $V_\lambda(G)$, and is called the {\em Weyl module} of $G$ indexed by 
$\lambda$. Symbolically:

\[\mbox{Irreducible representations of } G \stackrel{\mbox{Weyl}}{\Longleftrightarrow} \quad \mbox{partitions} \ \lambda. \]  

\[\mbox{Weyl module  } V_\lambda(G)  \longleftrightarrow \lambda. \] 
Weyl also proved that every finite dimensional representation of $G$ 
can be decomposed into irreducible representations--i.e.,
can be written as a direct sum of Weyl modules. Thus Weyl modules 
are the basic building blocks of the representation theory of 
$G$, and every finite dimensional representation of $G$ can be thought
of as a complex building made out of these blocks.

\begin{defn} (GCT2) \label{dgeoobs}
A (geometric) obstruction $O_{n,m}$ is a
Weyl module $V_\lambda(G)$ that lives on $X_{\#P}(n,m)$ but not
on $X_{NC}(n,m)$ (Figure~\ref{fig:class}).
\end{defn}

This means it  occurs as a subrepresentation of $G$ in
the space of polynomial\footnote{Well, these are not polynomial  in
the exact usual sense, but let us ignore that.}
 functions on $X_{\#P}(n, m)$ (which is a 
representation of $G$ since $G$ acts on $X_{\#P}(n,m)$) 
but not in the space of polynomial functions on $X_{NC}(n,m)$.
See  [GCT2] and \cite{GCTriemann} for a fully formal definition.

Existence of  an obstruction $O_{n,m}$, for every $n$, assuming
$m=2^{\log^a n}$, $a>1$ fixed, implies  that the inclusion (\ref{eqembed}) 
is not possible, since  $O_{n,m}$ cannot live  on $X_{NC}(n,m)$.
Thus  an obstruction blocks the inclusion (\ref{eqembed}). That is:

\begin{prop} (GCT2)
Existence of an obstruction $O_{n,m}$, for all $n\rightarrow \infty$, 
with $m=2^{\log^a n}$, $a>1$ fixed, implies $\perm(X)$, $\dim(X)=n$, 
cannot be represented linearly as 
$\det(Y)$, $\dim(Y)=m$. 
\end{prop}

See GCT1,2 for mathematical evidence and justification for why such
obstructions should exist.

This proposition leads to:

\begin{goal} \label{gobs} (GCT2)
Prove existence of such an
obstruction family $\{O_{n,m}=V_{\lambda_{n,m}}(G)\}$ using 
the  exceptional 
nature of $\perm(X)$ and $\det(Y)$, i.e., using the 
properties (P) and (D) in  Section~\ref{scharsym}.
\end{goal} 

This goal can be met unconditionally in the setting of the mathematical form
(Theorem~\ref{tmathform}):

\begin{theorem} \label{tmathobs}
There exists such an   obstruction family $\{O_{n,k}\}$ for the mathematical
form  of the $\#P \not = NC$ conjecture in characteristic zero 
(Theorem~\ref{tmathform}).
\end{theorem}

(There is no $m$ in this statement since Theorem~\ref{tmathform} is a
mathematical lower bound that does not depend on the complexity of the
polynomial in the traces in its statement.) 

This implies Theorem~\ref{tmathform}.
The notion of obstructions here is similar 
to the one in the general case. 

The proof of this result is  based on fundamental results of 
geometric invariant theory \cite{kempf,mumford}, 
the results of GCT1 and 2  based on it, and some representation theory 
\cite{ariki}. It is sketched in \cite{GCTriemann} with full details in GCT6.
It produces  a 
family $\{O_{n,k}=V_{\lambda_{n,k}}(G)\}$ of 
obstructions with a different $G = GL_n(\C) \times \cdots \times GL_n(\C)$ 
($2 k$ copies) than in the general complexity theoretic case.
Furthermore this family  is {\em strongly explicit} because 
the specification $\lambda_{n,k}$ of each $O_{n,k}$ has 
$O(n k)$ bitlength and  
can be constructed in $O(n k)$ time, in fact, in constant time 
by an $AC^0$  circuit (regardless of the 
complexity  of the polynomial in the traces in the statement of 
Theorem~\ref{tmathform}). This also provides a concrete evidence
that the geometric obstructions in GCT1 and 2 are fundamentally 
different from the trivial obstructions.

\section{C: Formalization of  the complexity barrier in  characteristic zero}
\label{sgct6}
We now proceed to describe the main result of GCT for the general
permanent vs. determinant problem in characteristic zero in the context
of  Goal~\ref{gobs}--this result  yields formalization of the
complexity barrier for this problem.

Towards that end, we  define some
representation-theoretic {\em stretching} functions. 
Let $F_{\lambda,n,m}(k)$ denote the number of copies of 
the Weyl module $V_{k\lambda}(G)$ that live on $X_{\#P}(n,m)$. 
Here $k \lambda$ denotes the partition obtained by multiplying each 
number in the integral sequence (partition) $\lambda$ by $k$.
Let $G_{\lambda,m}(k)$ denote the  dimension
of the subspace of invariants of $G_{det}$ in $V_{k \lambda}(G)$. Here
by an invariant we mean a fix point of $G_{\det}$. i.e., a point that
does not move under its action.

Given a polytope $Q$, we let $f_Q(k)$ denote the number of 
integer points in the dilated polytope $k Q$. 
More generally, given a parametrized  polytope $P(k)$
defined by a linear system of the form: 
\begin{equation} \label{eqpara}
P(k): \qquad A x \le k b + c,
\end{equation}
where $A$ is an $s \times t$ matrix, $x$ a variable $t$-vector, and 
$b$ and $c$ some constant $s$-vectors, let 
$f_P(k)$ denote the number of integer points in $P(k)$.

\begin{hypo} {\bf (PH1) [Positivity Hypothesis]} \label{hph} (GCT6)

\noindent (a) For every $\lambda,n,m\ge n$, 
there exists   an explicit  parametrized  polytope $P(k)=P_{\lambda,n,m}(k)$ 
such that 
\begin{equation} \label{eqP}
F_{\lambda,n,m}(k)=f_P(k).
\end{equation} 
If such a polytope exists it is guaranteed by a general result in GCT6 
that its dimension is $\poly(n)$ regardless of what $m$ is.
By explicit we mean 
it is given by 
a separation oracle \cite{lovasz} that, given any point $x$,
decides if $x \in P(k)$ and gives a separating hyperplane if it does not
in $\poly(n,\bitlength{m},\bitlength{x},\bitlength{k})$
time, where $\langle \ \rangle$ denotes the bitlength of
description.

\noindent (b) For every $m$, there exist an  explicit 
polytope  $Q=Q_{\lambda,m}$  such that
\begin{equation} \label{eqQ}
G_{\lambda,m}(k)=f_Q(k).
\end{equation}
If such a polytope exists it is guaranteed by a general result in GCT6 
that its dimension is $\poly(n)$ regardless of what $m$ is as long as
the length of $\lambda$ is $\poly(n)$ (as it will be in our applications).
Explicitness is defined similarly.
\end{hypo} 

Why PH1 should hold will be discussed in
Section~\ref{showph} below.
PH1, in particular, implies that $F_{\lambda,n,m}(k)$ and
$G_{\lambda,m}(k)$ have $\#P$ formulae. Positivity refers to the positive 
form of a $\#P$-formula, i.e., the absence of any negative sign
as in the usual formula for the permanent.

\begin{theorem} (GCT6) \label{tgct6}
There exists an explicit family $\{O_n\}$ of  obstructions 
for the permanent vs. determinant problem 
for $m=2^{\log^a n}$, $a>1$ fixed, $n\rightarrow \infty$, 
assuming, 

\begin{enumerate} 
\item PH1, and
\item OH (Obstruction Hypothesis): 

For all $n\rightarrow \infty$,  there exists $\lambda$ such that 
$P_{\lambda,n,m}(k) \not = \emptyset$ for all large enough $k$ and
$Q_{\lambda,m}=\emptyset$.
\end{enumerate}

In particular, this says that the complexity barrier for verification for the 
$\#P$ vs. $NC$ (i.e. permanent vs. determinant) 
problem in characteristic zero is crossed 
assuming  PH1. Or in other words, the complexity barrier for verification
is formalized with PH1 being the mathematical hypothesis 
involved in this formalization (cf. Section~\ref{sflip}).

The complexity barrier for discovery is also crossed assuming a
stronger positivity hypothesis PH4 that we do not specify here \footnote{The
hypothesis is called PH4 instead of PH2 because PH2 and PH3 are some other
positivity hypotheses related to PH1 that we do not discuss here}.
Or in other words, the complexity barrier for discovery and hence
the whole complexity barrier is also formalized, with PH1 and PH4 being
the mathematical hypotheses involved in this formalization.

Analogous results holds for the $P$ vs. $NP$ problem in characteristic zero.
\end{theorem}

For mathematical evidence and justification for why OH should hold
see [GCT2,6,7,8]. This result provides a formalization of the
complexity barrier in characteristic zero.
We  discuss its  reasonableness in
Section~\ref{showph} below. This addresses criterion C. 

The proof of Theorem~\ref{tgct6}, or rather the various results 
needed to justify its statement, is based on
the classical work of Hilbert in invariant theory \cite{mumford},
the fundamental work in algebraic geometry 
on the resolutions of singularities in characteristic zero
\cite{hironaka} and the 
cohomological works \cite{boutot,flenner} 
based on this resolution; cf. \cite{GCTriemann} and GCT6 for other references.

Here OH is a {\em complexity theoretic} hypothesis in the sense
that $m$ has to be small in comparison with $n$ for it to hold,
as in the original IOH for this problem (which is similar to the
IOH for  $P$ vs. $NP$ problem that we discussed before). 
In contrast, PH1 is a {\em mathematical} hypothesis in the sense
that there is no such restriction, and though there is a complexity 
theoretic component to it, it is not expected to be ``hard'' like the original 
$P \not = NP$ conjecture, because the dimensions of the various polytopes 
are already guaranteed to be $\poly(n)$, regardless of what $m$ is,
if they exist, and the complexity theoretic component is  basically 
linear programming on these small dimensional polytopes, which is 
generally regarded as  theoretically ``easy''.

Once  PH1 and PH4 are proved, the complexity
barrier is crossed and OH  becomes theoretically
feasible. Though, as we have already pointed out, actually proving OH
can be a mathematical challenge. 
Thus, in effect, this theorem has exchanged hard 
complexity theory (IOH) with hard mathematics (PH1+PH4+OH). What is
gained in this exchange is that the $P\not = NP$ conjecture 
should not stand in the way of this hard mathematics as it seemed to 
in the way of the original theoretically infeasible IOH.
The two concrete lower bounds of GCT that we discussed before 
illustrate that this exchange works in nontrivial instances.

As for as proving OH is concerned, there is nothing that we can
say at this point since it depends on the explicit forms of the polytopes 
in PH1/PH4. So we turn to PH1/4 next.

\section{Why should the positivity hypotheses 
hold and how to prove them?} \label{showph}
Reasonability of the formalization of the complexity barrier
in characteristic zero in Theorem~\ref{tgct6}
critically  hinges on  PH1 and PH4.
To show reasonability,
we have to  justify why these positivity hypotheses  should hold and 
why the $P\not = NP$ conjecture should not stand in the way of their proofs.
Let us just consider  PH1 (b), which is a basic prototype 
of the  general PH1 and PH4; for a discussion of the general PH1 and PH4,
see {GCT6,7,8].
So we have to now justify why PH1 (b)  should hold,  why the $P\not = NP$ 
conjecture should not stand in the way of its proof, and also briefly 
indicate GCT's plan for proving it.

The coefficients $G_{\lambda,m}(k)$ that occur in PH1 (b) 
are known as Kroncker coefficients in representation theory and hence we shall
refer to PH1 (b) as Kroncker PH1. It is a complexity 
theoretic version of the fundamental Kronecker problem in representation
theory that   has been intensively studied in the last century and 
is known to be formidable. And now 
it lies at the heart of this approach towards $P$ vs. $NP$. 

A special case of  Kronecker PH1 has been solved in representation theory.
for a very special class of Kronecker coefficients 
called Littlewood-Richardson (LR) coefficients. These 
are basically the multiplicities in the 
analogous  PH1 (b) that occurs in the proof of the
mathematical form of the $\#P\not = NC$ conjecture (Theorem~\ref{tmathform})
(formally LR coefficient is  the multiplicity of a given Weyl module
in the tensor product of other two given Weyl modules, 
but let us worry about its formal definition  here). 
We shall refer to this special case of the Kronecker PH1 
that occurs in the proof of the mathematical form as LR PH1.

Figure~\ref{fig:lift} gives  a pictorial
depiction of the plan to prove Kronecker PH1  
suggested  in [GCT4,6,7,8].
It strives  to extend the  proof of LR PH1 based 
on the theory of the  standard quantum group \cite{drinfeld,kashiwara,lusztig},
where a standard quantum group is  a mathematical object that plays
the same role in quantum mechanics that the classical group 
$G=GL_n(\C)$ plays in classical physics.  
There it comes out as a consequence of a  (proof of a) deep positivity result
\cite{kazhdan,lusztig}, which
we shall refer to as LR PH0. It says that every representation of 
the standard quantum group
has a canonical basis \cite{kashiwara,lusztig} 
whose structural coefficients that determine the action of the generators 
of the standard quantum group 
on this basis are polynomials in $q$ whose coefficients are 
all nonnegative  \cite{lusztig} (here $q$ stands for quantization 
parameter in physics). 
The only known proof of this positivity 
result is  based on the Riemann Hypothesis over finite fields 
proved in \cite{weil2}  and the related works \cite{beilinson}.
This Riemann Hypothesis over finite fields 
is  itself a deep positivity statement in mathematics,
from which   LR PH1 can  thus be 
deduced, as shown on the bottom row of Figure~\ref{fig:lift}.

If one were only interested in LR PH1, one  does not 
need this  powerful machinery, because it has a  much simpler 
algebraic  proof. 
But the plan to extend the
proof of LR PH1 to Kronecker PH1 and (its other needed generalizations)
in GCT4,6,7,8 is like a huge
inductive spiral. To  make it work,  one  needs a 
stronger inductive hypothesis than Kronecker PH1, more  precisely 
some  natural generalization of LR PH0 in this context which
we shall call Kronecker PH0.
Thus what is needed now 
is a systematic lifting  of the bottom row in Figure~\ref{fig:lift}
to the top, so as to complete the  commutative diagram, so to speak. 

\begin{sidewaysfigure} 
\begin{center} 
\epsfig{file=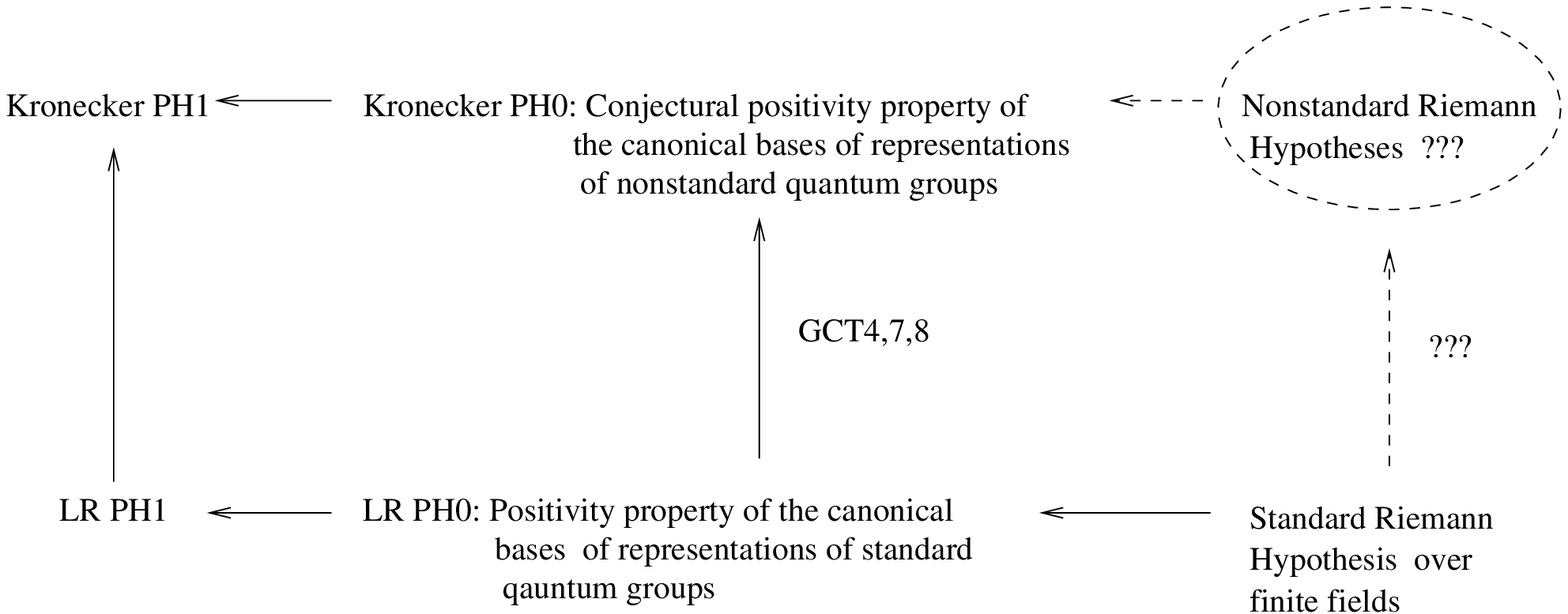, scale=.8}
\end{center}
      \caption{A commutative diagram}
      \label{fig:lift}
\end{sidewaysfigure}

Initial  steps in this direction have been taken in GCT4,7,8. 
First, GCT4 and 7 constructs a {\em nonstandard quantum group}, which
generalizes the notion of a standard quantum group \cite{drinfeld},
and plays the
same role in the context of Kronecker PH1
(and the   more general Plethysm PH1 which lies at the heart of 
the general PH1)
that the standard quantum group plays
in the context of  LR PH1. Second, GCT8 gives an algorithm to construct 
a canonical basis for a representation of the nonstandard quantum group
that is conjecturally correct and has the property Kronecker PH0,
which is some natural generalization of LR PH0, and is 
supported by experimental numerical evidence generated with the help
of a computer.

This numerical evidence is the main evidence for Kronecker PH0
and hence Kronecker  PH1 (i.e. PH1 (b)). 
It is reasonable because positivity (Kronecker PH0) is an extremely rigid 
property. Specifically, the structure coefficients involved in Kronecker 
PH0 are extremely complex mathematical quantities whose computation 
involves summing exponentially many terms of alternating signs.
If there were no reason for them to have any specific sign, 
they would be positive or negative with roughly the same probability,
naively speaking, and the probability that several hundred of these
mathematical quantities (which were computed) 
would all turn out to be of the same sign would be  absurdly  small 
(something like $2^{-500}$, again  naively speaking). 
Hence, this seems to be  a reasonble evidence  that PH1 (b)  should hold.

This numerical evidence for positivity (PH0) is in the spirit 
of the existing numerical evidence for  the Riemann Hypothesis,
to which PH0 is
intimately related as shown in Figure~\ref{fig:lift}. Existing 
numerical evidence for the  Riemann Hypothesis has been   obtained 
by  computing the first few zeros of the Riemann zeta function 
and then checking  if they really lie on the critical line. 
If they do (as they do so far) that gives
us a reasonable assurance  that the Riemann Hypothesis is true.
Especially so because the Riemann Hypothesis 
is an extremely rigid mathematical statement. By extremely rigid, we mean
if it were false we expect it to collapse completely and the zeros 
to get scattered  all over. Similarly PH0 is also expected to be 
an extremely rigid property since
it is intimately related to the Riemann Hypothesis. If were wrong, we again
expect it to collapse completely and the signs of the structural
coefficients  to be $+$ or $-$ with roughly equal probability.
That PH0 holds in the substantial numerical evidence obtained so far
gives us reasonable assuarance that it should hold just 
like the Riemann Hypothesis.

We also have to justify why the $P \not = NP$ conjecture should not
stand in the way of proving Kronecker PH0. 
This is because it is a natural mathematical  generalization of LR PH0, and
we know now that the $P\not = NP$ conjecture did not stand in the way of the
proof of LR PH0 in \cite{lusztig} and also in the way of the
proof of  the Riemanna Hypothesis over finite fields in \cite{weil2} 
on which the proof in \cite{lusztig} depends. The main point 
here is that however hard the Riemann Hypothesis and the related positivity
hypotheses in mathematics  may be, they are  not
not complexity theoretic hypotheses. That is, hard complexity theoretic 
statements have not been encoded in their statements. 
Of course, we cannnot guarantee that the $P\not = NP$ conjecture 
would not stand in the way of the proof of the usual Riemann Hypothesis.
But the fact that it did not stand in the way of the proof of 
its analogue over finite fields in \cite{weil2} gives us a reasonable 
assurance that it should not stand in the way of the proof of the usual
Riemann Hypothesis either. And similarly,  also in the 
the way of the proofs of the related positivity 
hypotheses in mathematics,  however hard they may be to prove.

Finally,  let us briefly discuss the plan for proving Kronecker PH0 proposed 
in [GCt7,8,11].
The plan is besically  to complete the 
commutative diagram in Figure~\ref{fig:lift}. What is needed 
for this completion  is an appropriate nonstandard 
extension of the Riemann hypothesis over finite fields and the related works
\cite{beilinson,weil2,kazhdan,lusztig}
from which  Kronecker PH0 can be deduced. This--the top-right corner 
of the diagram--is the   main  open problem at the heart of this approach. 
All the numerical evidence for the positivity suggests that 
what is needed in the top-right corner--some nonstandard extension
of the Riemann Hypothesis over finite fields--exists, but at present
we have no idea what it is. 
Whatever it is, the proof of PH0 promises to be a formidable affair.

To summarize, the discussion above suggests that there is now a good reason
to believe why the positivity hypotheses should hold, and why 
the $P\not = NP$ conjecture should not stand in the way of their
proofs--that is, why the formalization of the complexity barrier in
characteristic zero (Theorem~\ref{tgct6}) should be reasonable, and 
why this approach should thus penetrate the formidable circle around the
$P\not = NP$ conjecture in characteristic zero. But there is no reason
to believe that we are anywhere close to the proofs of these positivity
hypotheses. In fact, there is every reason to believe that we may be
very, very far away.

\section{C: Formalization of the complexity barrier  over finite fields} 
\label{sexpli}
Next  we briefly discuss  formalization of the complexity barrier
via the flip for the
usual $\#P$ vs. $NC$ and $P$ vs. $NP$ problems defined over finite
fields (e.g. boolean). 

This is based on the following.

\begin{conj}[Explicit proof conjecture] (GCT6) (Characteristic zero)
\label{cexpproof0} 
The geometric obstructions for the $\#P$ vs. $NC$ and $P$ vs. $NP$ problems 
in characteristic zero  defined in GCT1,2, and which 
were  discussed in Section~\ref{sgeomobs}, are extremely explicit. 
This means they are short, easy to verify, discover
and decode
(cf. Definitions~\ref{defnobstexplicit},\ref{defnobsstrongexp},\ref{ddecodeQ}).
Furthermore, short geometric obstructions
exist in the problems under consideration. Hence, extremely explicit  
proofs based on these geometric obstructions exist.
\end{conj}

(This is a strengthened form of PHflip in the earlier version of GCT6.)

\begin{conj} [Explicit proof conjecture] (GCT6) (Positive characteristic)
\label{cexpproofpos}
Analogous 
geometric obstructions for the $\#P$ vs. $NC$ and $P$ vs. $NP$ problems 
over algebraically closed fields of
positive characteristic  as defined in GCT6 are also extremely 
explicit. Here an algebraically closed field is assumed to be 
$\bar F_p$, the algebraic 
closure of the finite field $F_p$, for $p$ a large enough prime of $\poly(n)$
bitlength.
Furthermore, such short geometric obstructions exist.
Hence, extremely explicit proofs based on these geometric obstructions 
also exist.
\end{conj}

\begin{theorem}[GCT6] \label{texpproof}
An extremely explicit proof  for the $\#P$ vs. $NC$ or $P$ vs. $NP$ 
conjecture over $\bar F_p$ as in Conjecture~\ref{cexpproofpos} implies 
an extremely explicit proof for these problems over $F_p$, and 
also over the boolean field, i.e., for the usual forms of the nonuniform 
$\#P$ vs $NC$ and $P$ vs. $NP$ conjectures.
\end{theorem}

(Similar conjectures and results apply to other problems related to
the $P$ vs. $NP$ problem.)

This suggests the following strategy for proving the $\#P \not = NC$ 
and $P \not = NP$ conjectures:

\begin{enumerate} 
\item[I] Find  extremely explicit proofs  for the 
$\#P$ vs. $NC$ and $P$ vs. $NP$ problems
in  characteristic zero as in Conjecture~\ref{cexpproof0}.

\item[II] Lift them  to get  extremely explicit proofs for these problems over
$\bar F_p$, for a suitable $p$,  as in Conjecture~\ref{cexpproofpos}.

\item[III] By Theorem~\ref{texpproof} this implies extremely explicit proofs
over $F_p$ and the boolean field, the usual case.
\end{enumerate}

Theorem~\ref{texpproof}  reduces 
the search  for extremely explicit proofs over the  boolean field
(the usual case) to  the search for 
extremely explicit proofs over $\bar F_p$, for a large enough $p$.
The advantage of this reduction is that
the methods of algebraic geometry become applicable
over  algebraically closed fields. But algebraic geometry
in positive characteristic is harder than algebraic geometry in 
characteristic zero. This is why the goal is 
to get   an extremely explicit 
proof in characteristic zero  first (step I) and then lift it to 
algebraically closed fields of positive characteristic (step II). 
This {\em lift} 
promises to be a formidable mathematical challenge. A detailed
discussion of the problems  that need to be addressed in this lift
and additional relevant problems over finite fields will appear in GCT11.
Just as an example,
one of the crucial results that is used in the proof of Theorem~\ref{tgct6},
or rather a result needed to justify PH1 therein, is the 
resolution of singularities in characteristic zero \cite{hironaka}. Lifting
this resolution  to positive 
characteristic is one of the outstanding problems 
in algebraic geometry.  But the $P \not = NP$ conjecture is not expected 
to stand in the way of this lift, since the difficulty here is  
mathematical, not complexity theoretic. Similarly, 
if Conjecture~\ref{cexpproof0} can be proved, the 
$P \not = NP$ conjecture is not expected to stand in the way of
lifting its  proof (as in step II) to that of
Conjecture~\ref{cexpproofpos}.
Furthermore, the $P \not = NP$ conjecture is also not expected to stand in
the way of  decoding obstructions either, as we already discussed in
Section~\ref{sflip}.

Hence,  we will say that GCT is theoretically 
feasible over finite fields (the usual case) 
if it is   theoretical feasible 
in characteristic zero.  This definition of theoretical 
feasibility is different from the definition in Section~\ref{sflip}. We 
call the definition there  direct, and  the one here is indirect
(because it goes  via algebraically closed fields). 
This indirect definition of theoretical feasibility is also reasonable
since finally theoretical feasibility of an approach is only 
meant to say that the $P\not = NP$
conjecture should not stand in the way of the approach once it becomes 
theoretically feasible.

We have 
already  discussed in Section~\ref{showph}  why the formalization
of the complexity barrier in characteristic zero (Theorem~\ref{tgct6})
is reasonable--i.e., we gave 
a reasonble justification for theoretical feasibility
of  GCT in characteristic zero.
In conjunction with  Theorem~\ref{texpproof}
and the argument above, it can also be 
taken as  a reasonable justification  for its
theoretically feasibility in general--i.e., as a reasonable justification
for the formalization of the complexity barrier in general. 
Thus GCT meets  criterion C in general. 

But, as we also saw in Section~\ref{showph}, the proposed scheme for crossing
of the complexity barrier in characteristic zero
via the flip based on Theorem~\ref{tgct6}--i.e. for proving PH1/4--seems
to require nontrivial  extensions of profound  positivity results 
in several areas of mathematics,
such as algebraic geometry, representation theory, theory of 
quantum groups and so forth. Proposed scheme for crossing the complexity 
barrier in positive characteristic may need nontrivial extensions 
of positivity results 
in more related areas of mathematics such arithmetic algebraic geometry
and the theory of automorphic forms and so forth. 
Thus one can see that the implementation of this scheme for 
crossing the complexity barrier has turned out to be a highly nontrivial
affair.

\section{Is the complexity barrier really so formidable?} \label{sformidable}
That then leads to the final inevitable questions: is the complexity 
barrier really so formidable? Do we really need mathematics on this scale
to cross it?  There are two concrete  questions here.
First, is positivity  in abstract GCT necessary?
And second, is abstract GCT  necessary? (Recall that  by abstract GCT we mean 
any approach to  cross the complexity barrier via a flip 
based on extremely rigid explicit constructions driven by symmetries).

Let us deal with the first question first. Here the 
existing mathematical evidence and arguments [GCTflip] 
suggest that any abstract GCT approach 
to cross the complexity  barrier may  have to prove
some positivity results in the spirit of  PH1 and the  related 
positivity hypotheses either explicitly or {\em implicitly};
a detailed story and a precise meaning of the key 
phrase {\em implicit} will appear in the revised version of GCTflip.

Another evidence which  suggests that deep
positivity may be unavoidable in  rigid explicit constructions is provided 
by the fundamental work 
on explicit construction of Ramanujan graphs \cite{sarnak}. It may be 
conjectured that Ramanujan property is  rigid in the sense that very 
few graphs with bounded degree have this property,
though a random graph is close to being Ramanujan \cite{friedman}. 
In contrast, expansion property is not rigid because a random graph is an 
expander; i.e.  expansion is a naturalizable \cite{rudich} property.
Currently the only known proof of existence of Ramanujan graphs goes via
explicit construction, and its correctness is also based on a positivity
result--namely,
the Riemann hypothesis over finite fields (for curves). 
But Ramanujan property cannot be used to prove a lower bound.
Hence if a nontrivial  extension 
of the work surrounding this positivity result is needed for extremely 
rigid explicit constructions  that imply
hard lower bounds it should not be  too surprising.

All this suggests  positivity   lies at the heart 
of abstract  GCT and if we  follow this approach 
we have to be willing face deep positivity questions in mathematics.

That then leads to the second question: is abstract GCT necessary?
In this context
we have already remarked that there is no reason to believe that
abstract GCT is the only approach towards $P$ vs. $NP$, and that
there ought to be  nonconstructive approaches as well, perhaps several.
But these unnatural approaches may turn out to be even
harder than the natural approach. And in any case, 
at present we do not have a concrete 
alternative to GCT that meets the criteria A,B,C, and D.
\ignore{A concrete evidence towards this 
is provided in Figure~\ref{fig:summary}, which 
summarizes the discussion in this article.
Here it can be seen that as the lower bounds start moving towards the 
$P$ vs. $NP$ problem their existing
proof techniques  become more and more geometric,
explicit, and rigid for some  inexplicable reasons that we do not really
understand.}
In other words, we do not yet know how to approach 
the  $P$ vs. $NP$ problem  without  relying on extremely
rigid explicit constructions based on symmetries,
and we do not yet know how to avoid positivity in
carrying out such constructions.

We also have to keep in mind that
the complexity barrier is universal. So if deep
positivity is  needed to cross it   one way it may be 
be needed in one form  or the other to cross it   any other way.
Because that is what universal is supposed to mean.

All this  suggests that in the context of the $P$ vs. $NP$ problem
there may be after all something special about  geometry, symmetries,
extremely rigid explicit constructions, and positivity--the defining
features of abstract GCT--and unless and until we have a concrete 
alternative approach to cross the complexity barrier 
that meets the criteria A, B, C, and D, it may make sense
to just stick to this most  obvious and natural approach to cross the
complexity barrier towards the $P$ vs. $NP$ and related problems.

But paradoxically this most  obvious and natural approach 
has turned out to be such a massive affair that
it is fair to say  what has been done in GCT so far is only 
the easy initial part (formalization of the complexity barrier)--easy in 
comparison to what remains to be done (crossing)--and the 
hard part (positivity)  is yet to begin. And the hard part seems so 
hard that without an extensive collaboration between complexity theory
and mathematics further progress on this approach seems unlikely. It
is our hope that this small initial step towards the $P$ vs. $NP$  problem
will lead to such extensive collaboration between the two fields.

\end{document}